\documentclass[
  aps,
  prb,
  twoside,
  twocolumn,
  showpacs,
  floatfix,
  superscriptaddress,
  10pt,
  citeautoscript,
]{revtex4-1}

\usepackage{graphicx}
\usepackage{amsmath}
\usepackage{amssymb}
\usepackage{dcolumn}
\usepackage{units}

\newcommand{\eV}[0]{\text{eV}}
\newcommand{\meV}[0]{\text{meV}}
\newcommand{\cm}{\text{cm}}
\newcommand{\AAA}[0]{\text{\AA}}
\newcommand{\VBM}{\text{VBM}}

\newcommand{\La}{\text{La}}
\newcommand{\Br}{\text{Br}}
\newcommand{\Ce}{\text{Ce}}

\newcommand{\Sr}{\text{Sr}}

\newcommand{\Na}[0]{\text{Na}}

\newcommand{\Sect}[1]{Section~\ref{#1}}
\newcommand{\sect}[1]{Sect.~\ref{#1}}
\newcommand{\fig}[1]{Fig.~\ref{#1}}

\newcommand{\eq}[1]{Eq.~(\ref{#1})}

\newcommand{\tab}[1]{Table~\ref{#1}}

\newcommand{\mat}[1]{\ensuremath\mathbf{#1}}
\renewcommand{\vec}[1]{\ensuremath\boldsymbol{#1}}
\newcommand{\mydot}[0]{\bullet}

\newcommand{\myscale}{0.65}

\begin{document}


\title{
  A first-principles study of co-doping in lanthanum bromide
}

\author{Paul Erhart}
\email{erhart@chalmers.se}
\affiliation{Chalmers University of Technology, Department of Applied Physics, Gothenburg, Sweden}

\author{Babak Sadigh}
\affiliation{Physical and Life Sciences Directorate, Lawrence Livermore National Laboratory, Livermore, California 94550, USA}

\author{Andr{\'e} Schleife}
\affiliation{Department of Materials Science and Engineering, University of Illinois at Urbana-Champaign, Urbana, Illinois 61801, USA}

\author{Daniel {\AA}berg}
\email{aberg2@llnl.gov}
\affiliation{Physical and Life Sciences Directorate, Lawrence Livermore National Laboratory, Livermore, California 94550, USA}

\begin{abstract}
Co-doping of Ce-doped LaBr$_3$ with Ba, Ca, or Sr improves the energy resolution that can be achieved by radiation detectors based on these materials. Here, we present a mechanism that rationalizes of this enhancement that on the basis of first principles electronic structure calculations and point defect thermodynamics. It is shown that incorporation of Sr creates neutral $V_\Br-\Sr_\La$ complexes that can temporarily trap electrons. As a result, Auger quenching of free carriers is reduced, allowing for a more linear, albeit slower, scintillation light yield response. Experimental Stokes shifts can be related to different $\Ce_\La-\Sr_\La-V_\Br$ triple complex configurations. Co-doping with other alkaline as well as alkaline earth metals is considered as well. Alkaline elements are found to have extremely small solubilities on the order of 0.1 ppm and below at 1000\,K. Among the alkaline earth metals the lighter dopant atoms prefer interstitial-like positions and create strong scattering centers, which has a detrimental impact on carrier mobilities. Only the heavier alkaline earth elements (Ca, Sr, Ba) combine matching ionic radii with sufficiently high solubilities. This provides a rationale for the experimental finding that improved scintillator performance is exclusively achieved using Sr, Ca, or Ba.
The present mechanism demonstrates that co-doping of wide gap materials can provide an efficient means for managing charge carrier populations under out-of-equilibrium conditions. In the present case dopants are introduced that manipulate not only the concentrations but the electronic properties of {\em intrinsic} defects without introducing additional gap levels. This leads to the availability of shallow electron traps that can temporarily localize charge carriers, effectively deactivating carrier-carrier recombination channels. The principles of this mechanism are therefore not specific to the material considered here but can be adapted for controlling charge carrier populations and recombination in other wide gap materials.
\end{abstract}
 
\maketitle

\section{Introduction}
\label{sect:introduction}

Many applications in nuclear and radiological surveillance, high-energy physics, and medical imaging rely on scintillator materials \cite{Rod97, Kno10}, which enable the energy resolved detection of high energy radiation \cite{NelGosKna11}. 
The energy resolution that can be accomplished increases with luminosity, which is usually related to the efficiency of the process by which the energy of incoming radiation quanta (typically in the keV to MeV range) is converted to lower energy photons (on the order of a few eV). The achievable resolution is, however, further limited by the non-linearity of the scintillation response to the energy of the incident radiation \cite{Dor10}, which arises from the competition between non-radiative quenching, defect carrier trapping, as well as activator capture and subsequent emission \cite{Dor05, Vas08, KerRosCan09, BizMosSin09, PayMosShe11}.

Recent work showed that the energy resolution of Ce-doped LaBr$_3$ can be significantly improved by co-doping. The concept was first realized experimentally by Yang {\it et al.} for samples co-doped with Sr \cite{YanMenBuz12}. Later Alekhin {\it et al.} achieved an improvement of energy resolution from 2.7 to 2.0\%\ at 662\,keV using Sr and Ca \cite{AleHaaKho13}. A more comprehensive investigation including both the alkaline as well as alkaline earth series revealed that better performance is only achieved when using the heavier elements of the latter series (Sr, Ca, Ba) \cite{DorAleKho13}. To explain these observations it has been suggested that doping with Sr, Ca, or Ba causes \cite{AleWebKra14}
({\it i}) a reduction of the non-radiative recombination rate,
({\it ii}) an increase of the so-called escape rate of the carriers from the quenching region, or
({\it iii}) an increase in the trapping rate of $\Ce^{3+}$.
The experimental investigations also revealed three distinct optical signatures associated with Ce, which have been interpreted as evidence for the presence of three different Ce environments in the co-doped material. In contrast, only one such feature can be identified in LaBr$_3$:Ce. Later, it was argued by the present authors that Sr-doping causes the creation of shallow electron trap complexes, which leads to reduced Auger quenching \cite{AbeSadSch14}.

The present paper describes the argumentation in detail and presents a careful analysis of the thermodynamic properties and electronic structures of the most important intrinsic and extrinsic defects -- including their complexes -- in Ce and Sr-doped LaBr$_3$ including self-trapped polaronic configurations. To demonstrate that the present model is consistent with experimental observations, a comprehensive set of calculated absorption and emission energies for the relevant Ce complexes is carried out. We obtain very good overall agreement, which enables us to correlate optical signatures with individual defect configurations.

Finally, it is shown that the solubilities of Sr, Ca, and Ba near the synthesis temperature are several 100 ppm, whereas much lower values are obtained for the alkaline metals. Be and Mg yield large solubilities but also cause large lattice distortions and effectively act as interstitials. These findings provide a rationale for why only Sr, Ca, and Ba have been experimentally found to improve scintillation performance.

As part of the present work we also introduce a convenient method for generating supercells with optimal shapes for defect calculations and finite-size scaling. The approach described in the appendix is applicable to arbitrary lattice types.

The paper is organized as follows. \Sect{sect:methodology} describes the computational methodology used in this work. The results regarding the thermodynamic properties of intrinsic defects as well as Sr and Ce dopants are presented in \sect{sect:results}. Stokes shifts for single substitutional Ce and the triple complexes are reviewed and compared to experimental data and a dopant solubility analysis are presented. To conclude, we discuss our findings in light of scintillator performance in \sect{sect:discussion}. The Appendices provide additional information concerning defect thermodynamics, solubilities, and finite size scaling of formation energies.

\section{Methodology}
\label{sect:methodology}

\subsection{General computational parameters}

Calculations were performed within density functional theory (DFT) using the projector augmented wave (PAW) method \cite{Blo94, *KreJou99} as implemented in the Vienna ab-initio simulation package \cite{KreHaf93, *KreHaf94, *KreFur96a, *KreFur96b}. Exchange-correlation (XC) effects were treated within the generalized gradient spin approximation \cite{PerBurErn96}. DFT+$U$ type on-site potentials \cite{DudBotSav98} were included for both La-$4f$ ($U_{\text{eff}}=10.3\,\eV$) and Ce-$4f$ states ($U_{\text{eff}}=1.2\,\eV$) in order to obtain the correct ordering of La-$5d$ and $4f$ states and to reproduce experimental Ce-$4f$ ionization energies \cite{CzySaw94, CanChaBou11, AbeSadErh12}. The plane wave energy cutoff was set to 230\,eV and Gaussian smearing with a width of 0.1\,eV was used to determine occupation numbers.

Lanthanum bromide adopts a hexagonal lattice structure in space group 176 (P$6_3$/m) with La and Br ions occupying Wyckoff sites $2c$ and $6h$, respectively. Using the aforementioned computational parameters one obtains lattice parameters of $a=8.140\,\AAA$ and $c=4.565\,\AAA$ to be compared with experimental values of $a=7.9648(5)\,\AAA$ and $c=4.5119(5)\,\AAA$, respectively \cite{KraSchSch89}.

Spin-orbit interaction was not included self-consistently but rather added as a perturbation to the $4f$ states for optical transitions according to 
\begin{align*}
 \Delta E_{\text{so}} = 
 \begin{cases}
   -2\xi_{4f} & j =  \nicefrac{5}{2} \\
   \nicefrac{3}{2}\xi_{4f} & j = \nicefrac{7}{2},
 \end{cases}
\end{align*}
where $\xi_{4f}=0.1\,\eV$ as obtained from the $4f$ splitting in a Ce-$4f^05d^1$ configuration.

\subsection{Excited states}

Excited Ce-$4f^0$ states were obtained in a similar approach as used by Canning and co-workers. \cite{CanChaBou11,ChaBouCho14}. First, a subspace of Ce-$4f$ states is identified by selecting states whose projection onto spherical harmonics with $l=3$ exceeds a suitably chosen threshold. This is made possible by the localized nature of the rare-earth $4f$ states. In this particular case, a Ce $4f^15d^0\rightarrow 4f^0 5d^1$ excitation was emulated by completely deoccupying the $4f$ manifold while keeping the number of electrons fixed. This results in a narrow, almost degenerate, group of unoccupied group of 14 $4f$ bands inside the fundamental electronic gap. The occupation numbers of Kohn-Sham states above this group is then occupied using Gaussian smearing, allowing for excited state structural relaxation.

\subsection{Defect calculations}

Defect formation energies were calculated using a well established thermodynamic formalism \cite{ZhaNor91, *ErhAbeLor10} that is summarized for convenience in Appendix~\ref{sect:eform}. Lanthanum bromide has a small dielectric constant and some defects adopt large charge states ($|q|\leq 4$). As a result, image charge interactions are substantial and care must be taken to remove finite-size effects in order to obtain formation energies for the dilute limit \cite{LanZun08}. To this end a careful finite-size scaling study was carried out that is summarized in Appendix~\ref{sect:scaling}. This Appendix also introduces a general and convenient method for constructing suitable supercells with optimal shape given a certain system size, which allows one to obtain a dense sampling of different system sizes also in the case of low crystal symmetry. Hole polarons were studied using the polaron self-interaction correction method (pSIC) \cite{SadErhAbe14}.

The bulk of the data presented in the following were obtained using 168-atom supercells. $\Gamma$-point sampling was found to be sufficient to converge defect formation energies to better than 0.05\,eV. Configurations were relaxed until ionic forces were less than 10\,\meV/\AA. Potential alignment as well as periodic image charge corrections were taken into account to correct for finite size effects as detailed in Appendix~\ref{sect:scaling} \cite{LanZun08}. 

It is well known that conventional XC functionals including the ones used in the present work commonly lead to an underestimation of band gaps. As discussed for example in Refs.~\onlinecite{PerZhaLan05, ErhAlb07, ErhAlb08, LanZun08, ErhAbeLor10}, this error also affects defect formation energies and therefore also defect concentrations. Hybrid XC functionals, which combine conventional DFT functionals with Fock exchange, are often found to improve the band gap and are therefore expected to also yield improved formation energies. For the present case, however, hybrid functionals are ill-suited since they cannot even qualitatively describe the position of the occupied Ce-$4f$ level in LaBr$_3$. This is related to the distinct character of the electronic states involved. They are associated with widely different levels of localization and thus the effective screening cannot be parametrized using a single (static) mixing parameter.

In the present study we therefore resort to a simple correction scheme \cite{PerZhaLan05, ErhAlb07, ErhAlb08, LanZun08} that shifts the formation energies based on the offset between the ``true'' band edges and the ones obtained within the underlying computational framework, i.e., in the present case DFT+$U$ calculations. The offsets were determined using the $G_0W_0$ method \cite{Hed65, HedLun70}, which was previously found to yield a much improved description of the band structure of LaBr$_3$ compared to both DFT and DFT+$U$ \cite{AbeSadErh12}. Calculations for the primitive cell were carried out using a $\Gamma$-centered $3\times3\times6$ $\vec{k}$-point mesh and PAW data sets optimized for $GW$ calculations, which require also unoccupied higher energy states to be well described. The dielectric tensor was computed for energies up to 200\,eV above the CBM, equivalent to 1024 unoccupied bands. The offsets obtained in this way are $\Delta E_\text{VB}=-1.2\,\eV$ for the valence band edge and $\Delta E_\text{CB} = +0.5\,\eV$ for the conduction band edge. This increases the DFT+$U$ band gap from 3.6\,eV to 5.3\,eV, which is in much better agreement with the experimental value of 5.9\,eV \cite{DorLoeVin06}. The correction scheme was only applied to defect charge states that did not include occupied localized states inside the band gap. As this distinction can be ambiguous, additional $G_0W_0$ calculations were carried out for 96-atom supercells of the most important defect configurations to verify the results from the correction scheme. In these calculations the Brillouin zone was sampled using the $\Gamma$-point only and the same $GW$ optimized PAW data sets as before were employed. The dielectric function was calculated up to 36\,eV above the CBM. Based on this comparison the error in the formation energies is estimated to be 0.2\,eV or less whereas for the transition levels the error is estimated to be below 0.1\,eV.

Finally, defect concentrations were obtained using the calculated formation energies on the basis of a self-consistent solution of the charge neutrality condition, which has been described in detail in Refs.~\onlinecite{ErhAlb08, ErhAbeLor10}.

\section{Results}
\label{sect:results}

\subsection{Intrinsic defects}

\begin{figure*}
  \centering
\includegraphics[scale=\myscale]{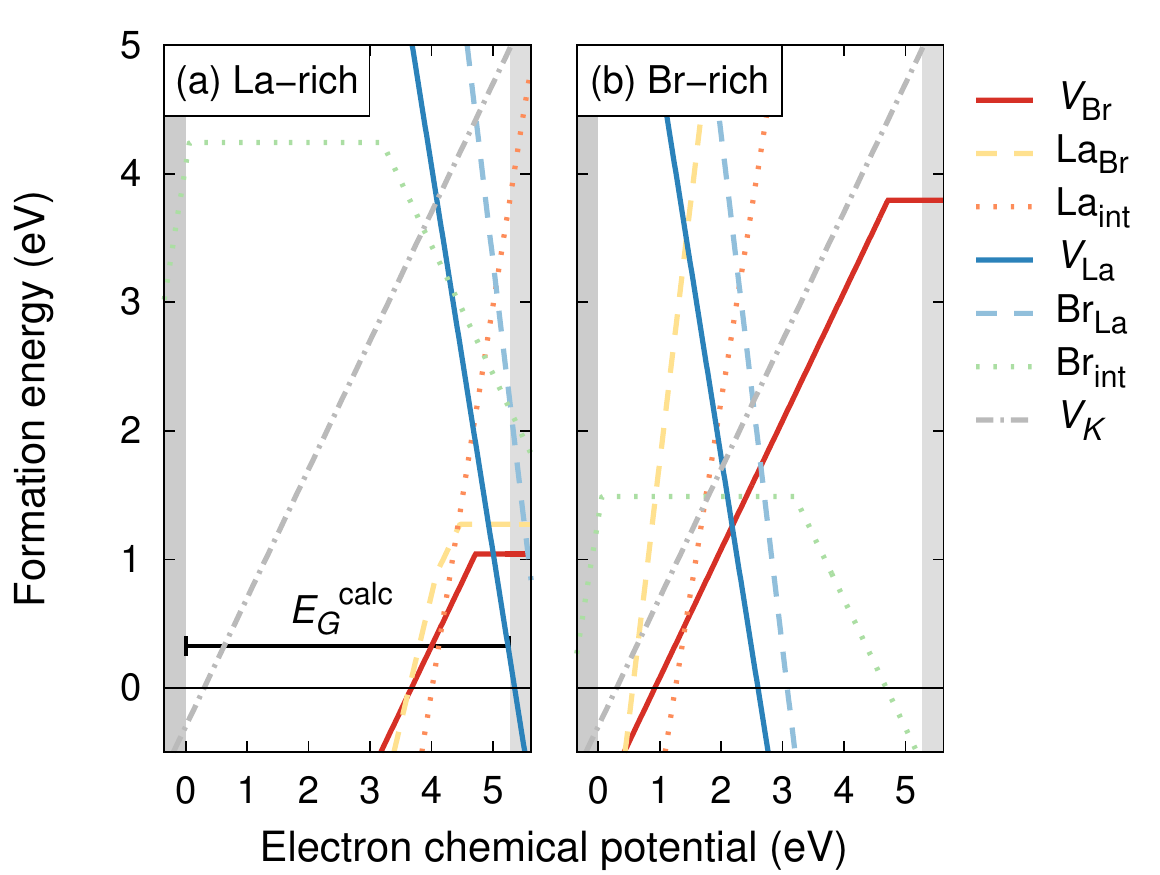}
\includegraphics[scale=\myscale]{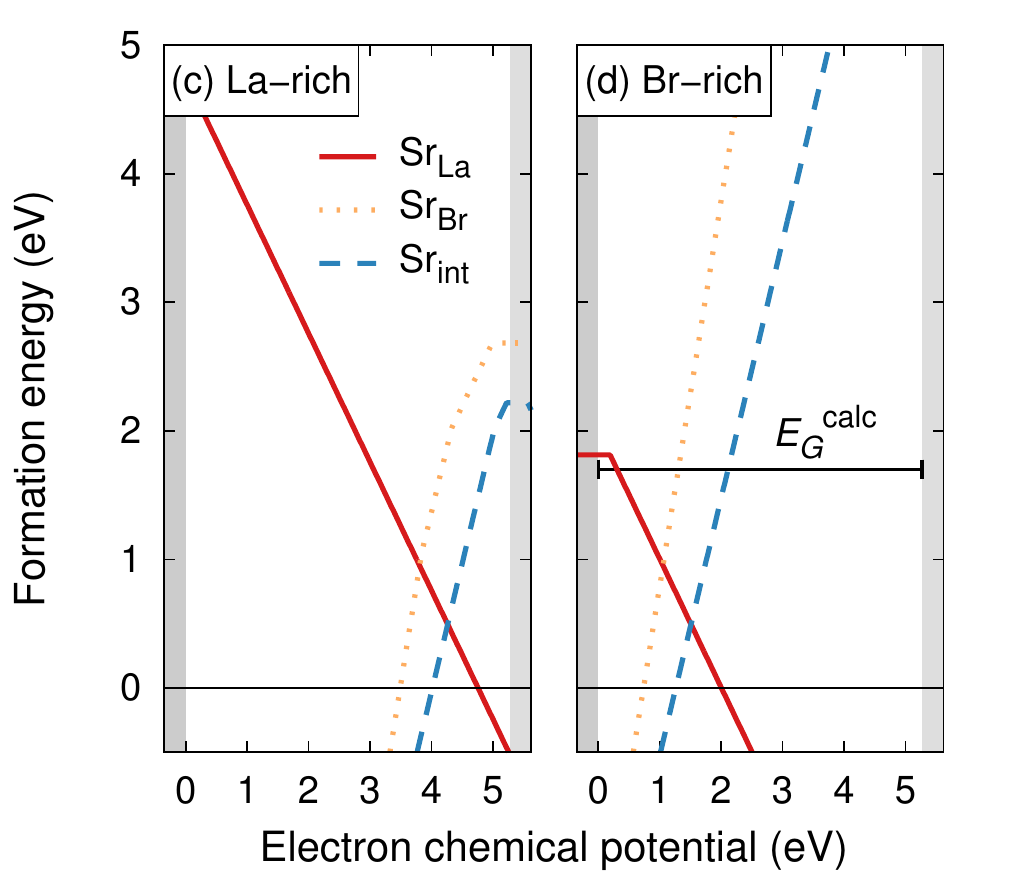}
  \caption{
    Defect formation energies for (a,b) intrinsic defects including $V_K$-centers as well as (c,d) Sr.
    Formation energies for Sr related defects where computed assuming equilibrium with SrBr$_2$ (compare \sect{sect:solubilities}).
  }
  \label{fig:eform}
\end{figure*}

\begin{figure*}
  \centering
\includegraphics[scale=\myscale]{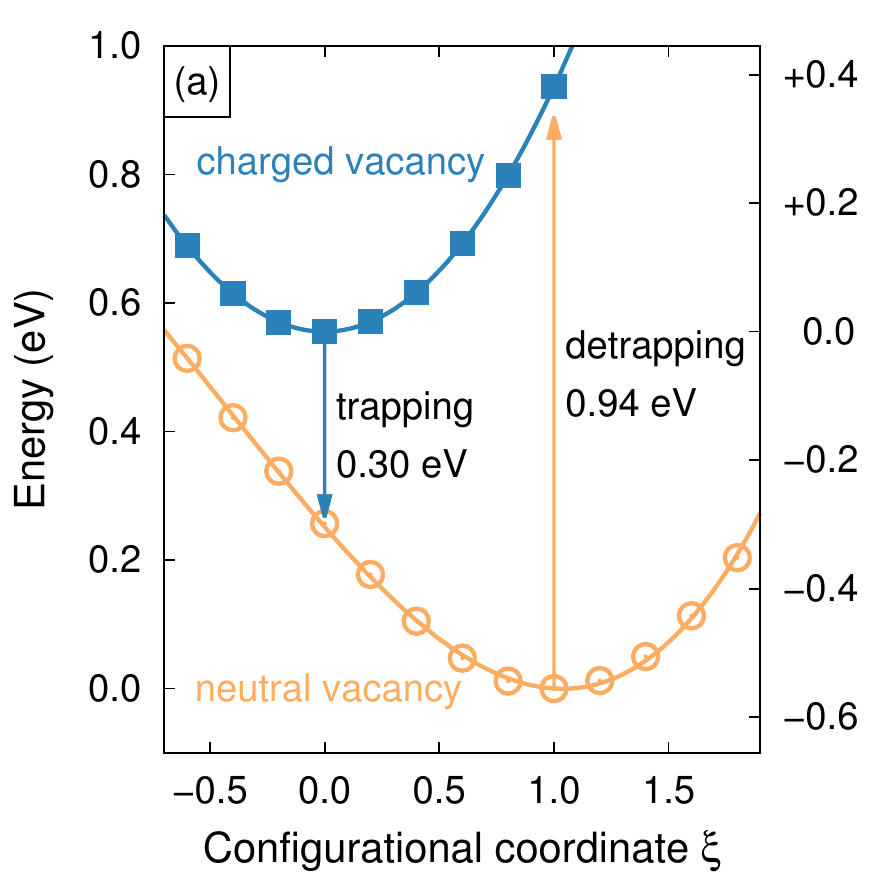}\hspace{12pt}
\includegraphics[scale=\myscale]{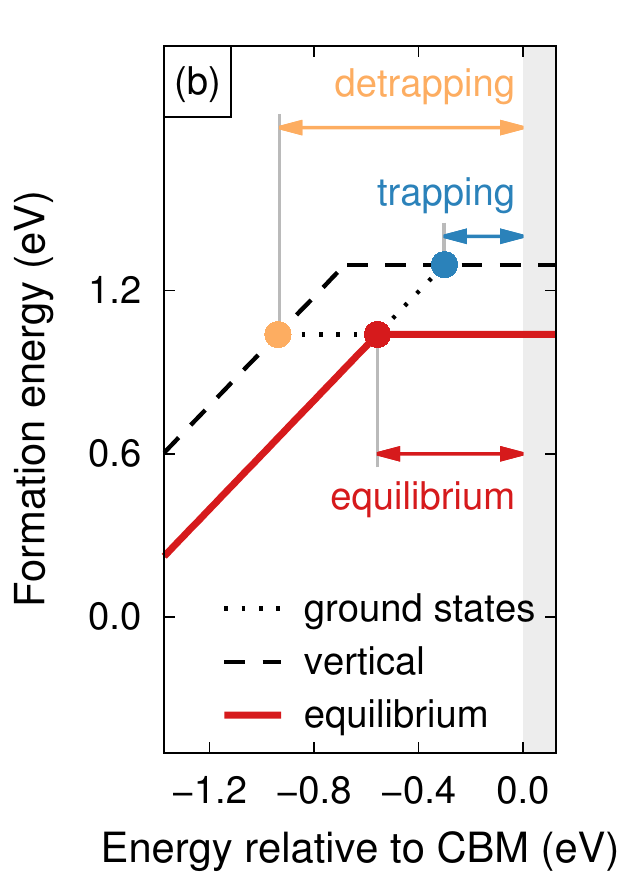}
\includegraphics[scale=0.07]{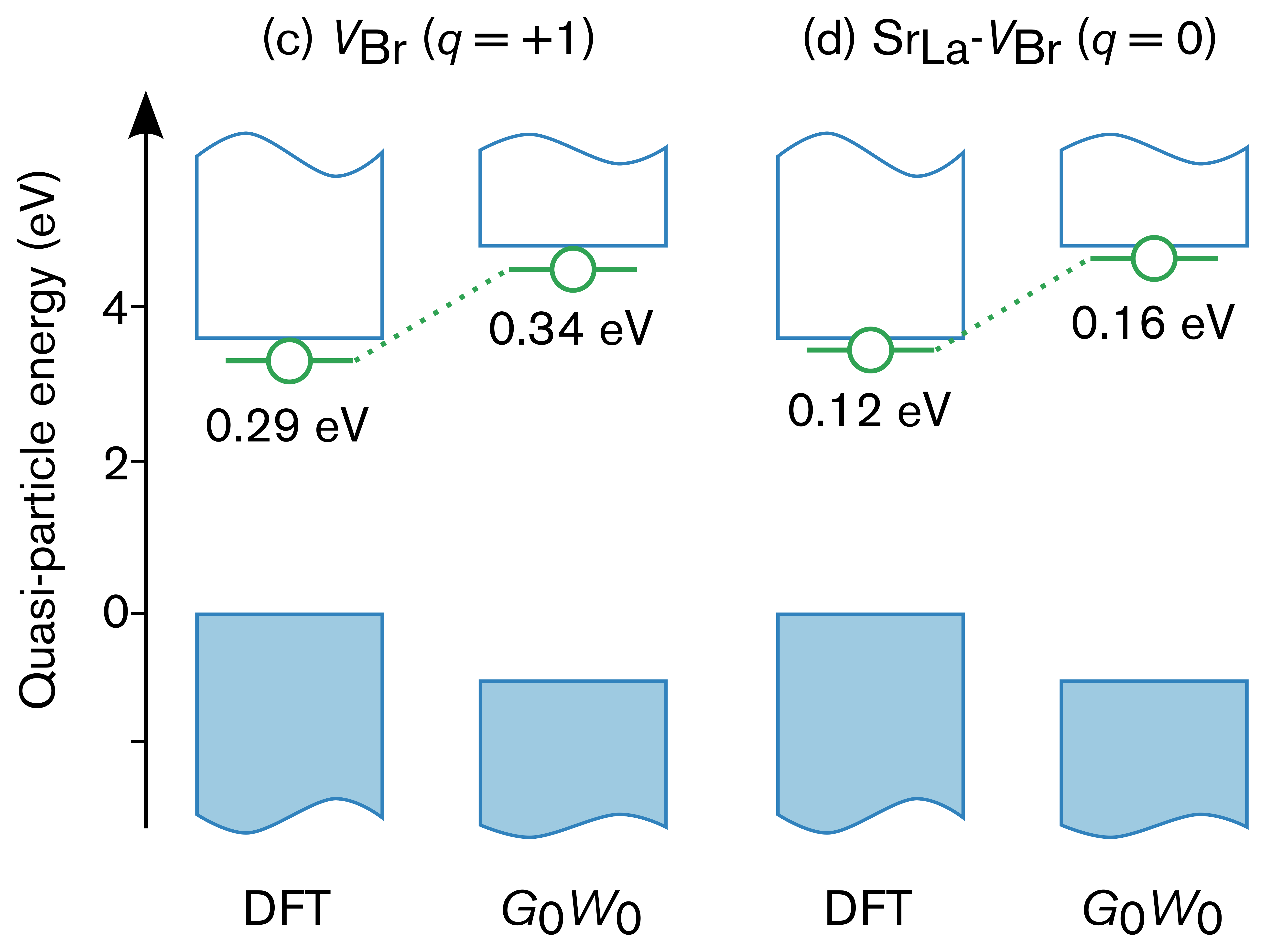}
  \caption{
    (a) Configuration coordinate diagram illustrating the relation between neutral and charged Br vacancies. For convenience the electron chemical potential was assumed to coincide with the CBM ($\mu_e=\epsilon_\text{CBM}$) such that the energy differences indicated by the vertical arrows correspond directly to the transition levels indicated in the middle panel.
    (b) Formation energy of Br vacancy configurations for La-rich conditions as a function of the electron chemical potential in the vicinity of the conduction band edge, compare \fig{fig:eform}(a). The solid and dotted lines are based on fully relaxed configurations, corresponding to minima in panel (a). Dashed lines indicate formation energies representing vertical transitions in panel (a). The horizontal blue and orange arrows thus correspond to the vertical arrows in panel (a). The red horizontal arrow is equivalent to the energy difference between the minima in panel (a).
    (c,d) Quasi-particle spectra from DFT and $G_0W_0$ calculations for both (c) the isolated Br vacancy and (d) a $\Sr_\La-V_\Br$ complex. The position of the defect level relative to the CBM changes by less than 0.1\,eV when comparing the two types of calculations. Note that in both cases the defect level is unoccupied in equilibrium.
  }
  \label{fig:confcoord}
\end{figure*}

\begin{figure}
  \centering
\includegraphics[scale=\myscale]{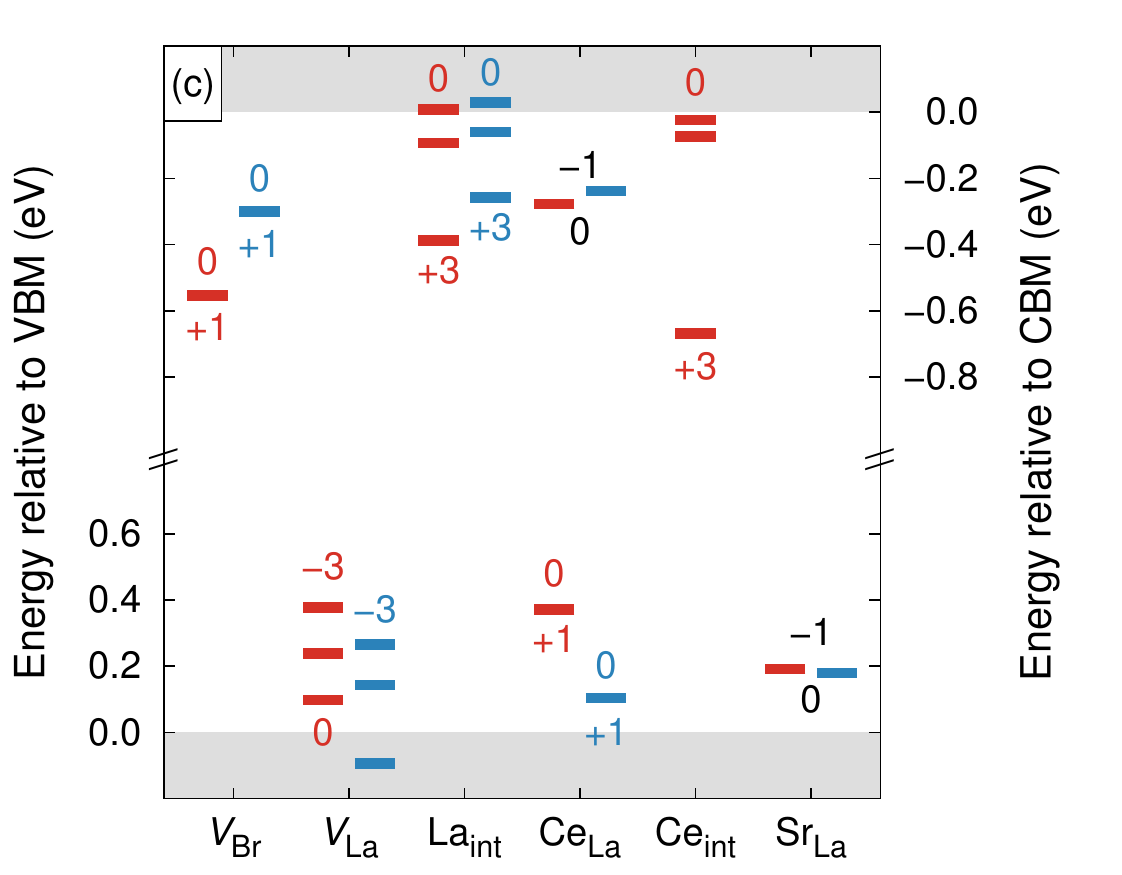}
  \caption{
    Equilibrium transition in red (left hand columns) and trapping levels in blue (right hand columns) obtained in the fashion indicated in \fig{fig:confcoord}.
  }
  \label{fig:levels}
\end{figure}

\begin{figure}
  \centering
\includegraphics[scale=0.19]{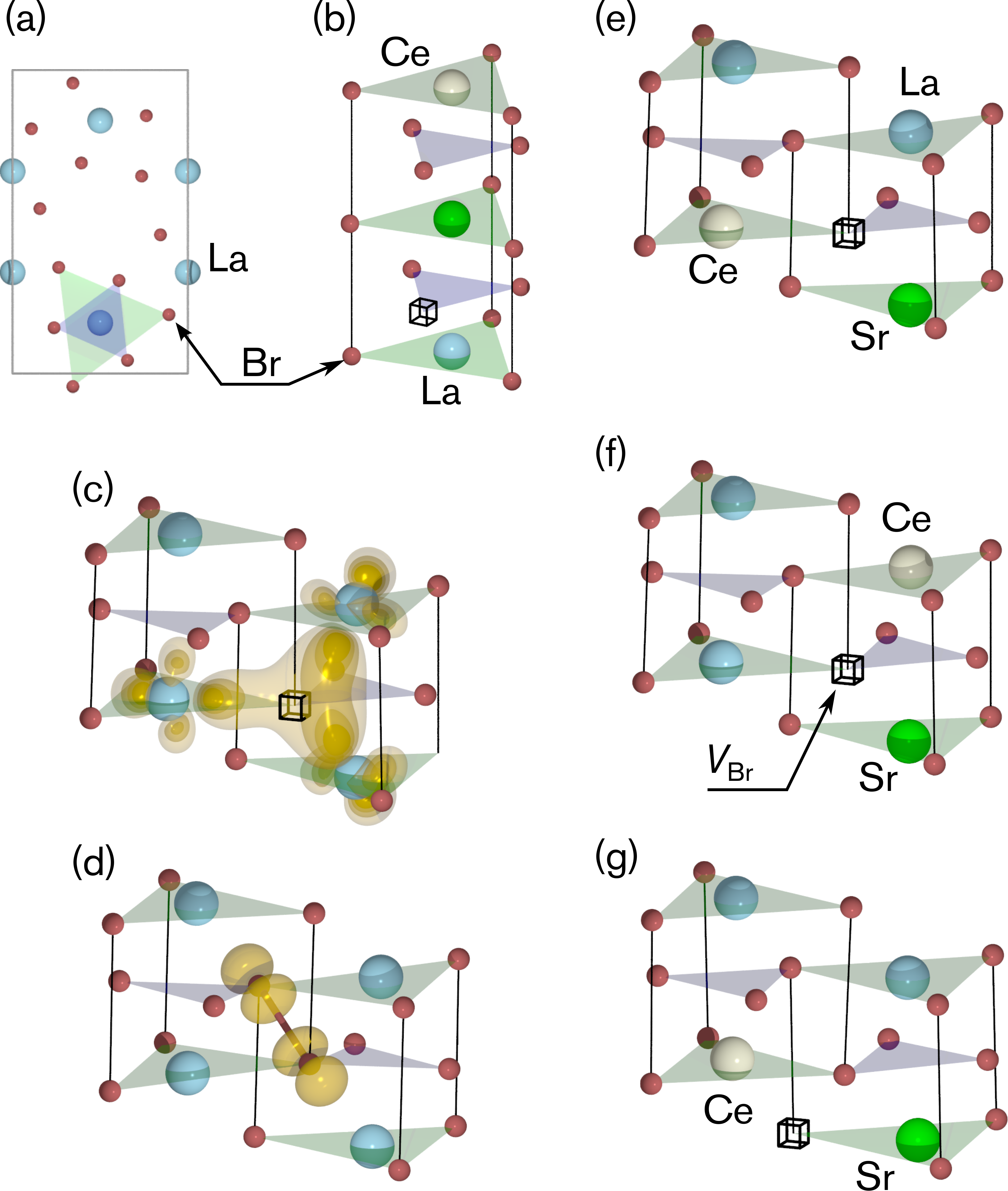}
  \caption{
    Structure of lanthanum bromide and defect configurations therein.
    (a) Projection onto the (0001) basal plane. Large blue and small red spheres represent La and Br sites, respectively. The shaded triangles correspond to atomic layers containing both atom types (green) and Br ions only (lila), respectively.
    (b) Perspective view. A representative Br vacancy site is indicated by the cube and Sr as well as Ce substitutional sites are shown by large green and light gray spheres, respectively.
    (c) Isosurface of localized vacancy level illustrating that it is composed of $5d$ states derived from the three nearest cations. 
    (d) Isosurface of localized hole level for lowest energy $V_K$-center 
    (e--g) Configurations of the $\Ce_\La-\Sr_\La-V_\Br$ triple complex, in which both Sr and Ce are  within the first neighbor shell of the Br vacancy.
    (e) in-plane $\Ce_\La-V_\Br$ and out-of-plane $\Sr_\La-V_\Br$.
    (f) out-of-plane $\Ce_\La-V_\Br$ and out-of-plane $\Sr_\La-V_\Br$.
    (g) out-of-plane $\Ce_\La-V_\Br$ and in-plane $\Sr_\La-V_\Br$.
  }
  \label{fig:structures}
\end{figure}

In this section the defect equilibria in nominally pure LaBr$_3$ are investigated. To this end, one first requires knowledge of the formation energies of intrinsic defects, which are shown for La and Br-rich conditions (see Appendix~\ref{sect:eform}) in \fig{fig:eform}(a,b). It is apparent that under both La and Br-rich conditions the most important intrinsic donor and acceptor defects are Br and La vacancies, respectively, with interstitials and antisites playing minor roles.

Given the respective formation energies it is straightforward to compute the equilibrium transition level between charge states $q_1$ and $q_2$ according to
\begin{align}
  \epsilon_{eq}(q_1/q_2) &= \frac{\Delta E_f(q_1) - \Delta E_f(q_2)}{q_2-q_1}.
  \label{eq:trans}
\end{align}
Experimentally, these can be detected for example by deep level transient spectroscopy. The levels calculated according to \eq{eq:trans} are shown as red bars (left hand columns) in \fig{fig:levels}, which reveals that both types of vacancies are associated with deep equilibrium transition levels. Note in particular that the +1/0 equilibrium transition level of the Br vacancy is located 0.55\,eV below the CBM. In $G_0W_0$ calculations, the quasi-particle energies associated with $V_\Br$ defect levels shift by less than 0.1\,eV relative to the CBM, which provides strong support for this positioning of the defect level with respect to the conduction band edge. This is illustrated in \fig{fig:confcoord}(c,d), which compares quasi-particle energies from DFT and $G_0W_0$ calculations for both isolated and complexes Br vacancies. Note that for consistency, only transition levels obtained from DFT+$U$ calculations including band gap corrections are reported from here on.

Deep defects are typically associated with pronounced changes in the ionic positions between different charge states. This is indeed shown to be the case for Br vacancies in \fig{fig:confcoord}(a), which shows the potential energy surfaces (PES) for both neutral and charged vacancies along the configuration coordinate connecting the respective minima. A particular configuration is given in terms of the configuration coordinate $\xi$ as
\begin{align}
  \vec{R} &= \vec{R}_{min}^{+1} + \xi \left( \vec{R}_{min}^0 - \vec{R}_{min}^{+1} \right) \big/ a_{\text{FC}},
  \label{eq:confcoord}
\end{align}
where $\vec{R}_{min}^q$ denotes the minimum of the PES for charge state $q$ and $a_{\text{FC}}=|\vec{R}_{min}^0 - \vec{R}_{min}^{+1}|$ measures the structural difference between the two geometries.

In the ideal structure, Br sites are surrounded by three cations (two out-of-plane and one in-plane with respect to the \{0001\} basal plane) at distances between 3.1 and 3.2\,\AA{} and eight Br ions (two in-plane, six out-of-plane) at distances of 3.6--3.7\,\AA, see \fig{fig:structures}. In the charged vacancy configuration, the La neighbors of the vacant site move outward by 0.2--0.3\,\AA{} while the Br neighbors move inward by up to 0.3\,\AA. These relaxations are inverted for the neutral vacancy as La and Br neighbors are shifted inward and outward, respectively compared to the ideal structure. The large differences in ionic configuration are, as shown in \fig{fig:confcoord}(a), associated with substantial relaxation energies of 0.25 and 0.38\,eV on the neutral and charged PES, respectively. As illustrated in \fig{fig:structures}(c), the defect level is predominantly composed of $5d$ states localized at the three cations surrounding the vacant Br site, which as discussed below is of crucial importance for understanding the effect of Br vacancies on the optical signature of Ce. The defect level can act as an efficient electron trap, effectively removing carriers from the light-generation process during the instrumentation pulse shape-time.

In terms of electronic trapping, also ``vertical'' transitions are important, which are indicated by the vertical arrows in \fig{fig:confcoord}(a). They can be calculated in a fashion similar to \eq{eq:trans} but imposing the constraint of fixed ionic position. Specifically, the trapping level for charge state $q_1$ is obtained as
\begin{align}
  \epsilon_{tr}(q_1/q_2) &= \frac{\Delta E_f(q_1;\vec{R}_{min}^{q_1}) - \Delta E_f(q_2;\vec{R}_{min}^{q_1})}{q_2-q_1},
\end{align}
where $\Delta E_f(q_2;\vec{R}_{min}^{q_1})$ is the formation energy computed for charge state $q_2$ at the ionic coordinates corresponding to the equilibrium positions in charge state $q_1$. All trapping levels presented in this paper were calculated with respect to the respectively more favorable charge state, e.g., $q=+1\rightarrow 0$ for the Br vacancy and $-3\rightarrow-2$, $-2\rightarrow-1$ \ldots for the La vacancy.

Equilibrium transition and trapping levels can also be identified with crossing points in plots of the formation energy {\it vs.} electron chemical potential as illustrated in \fig{fig:confcoord}(b). In this figure, dashed lines correspond to formation energies computed in the geometry of the respective other charge state, i.e. in the case of the neutral vacancy at $\vec{R}_{min}^{+1}$ and vice versa. The equivalence of trapping and detrapping type transitions relative to the charge vacancy are shown by the blue and orange arrows in Figs.~\ref{fig:confcoord}(a) and (b). Both equilibrium transition and trapping levels are compiled in \fig{fig:levels} for the most relevant intrinsic and extrinsic (see below) defects.

\begin{figure*}
  \centering
\includegraphics[scale=\myscale]{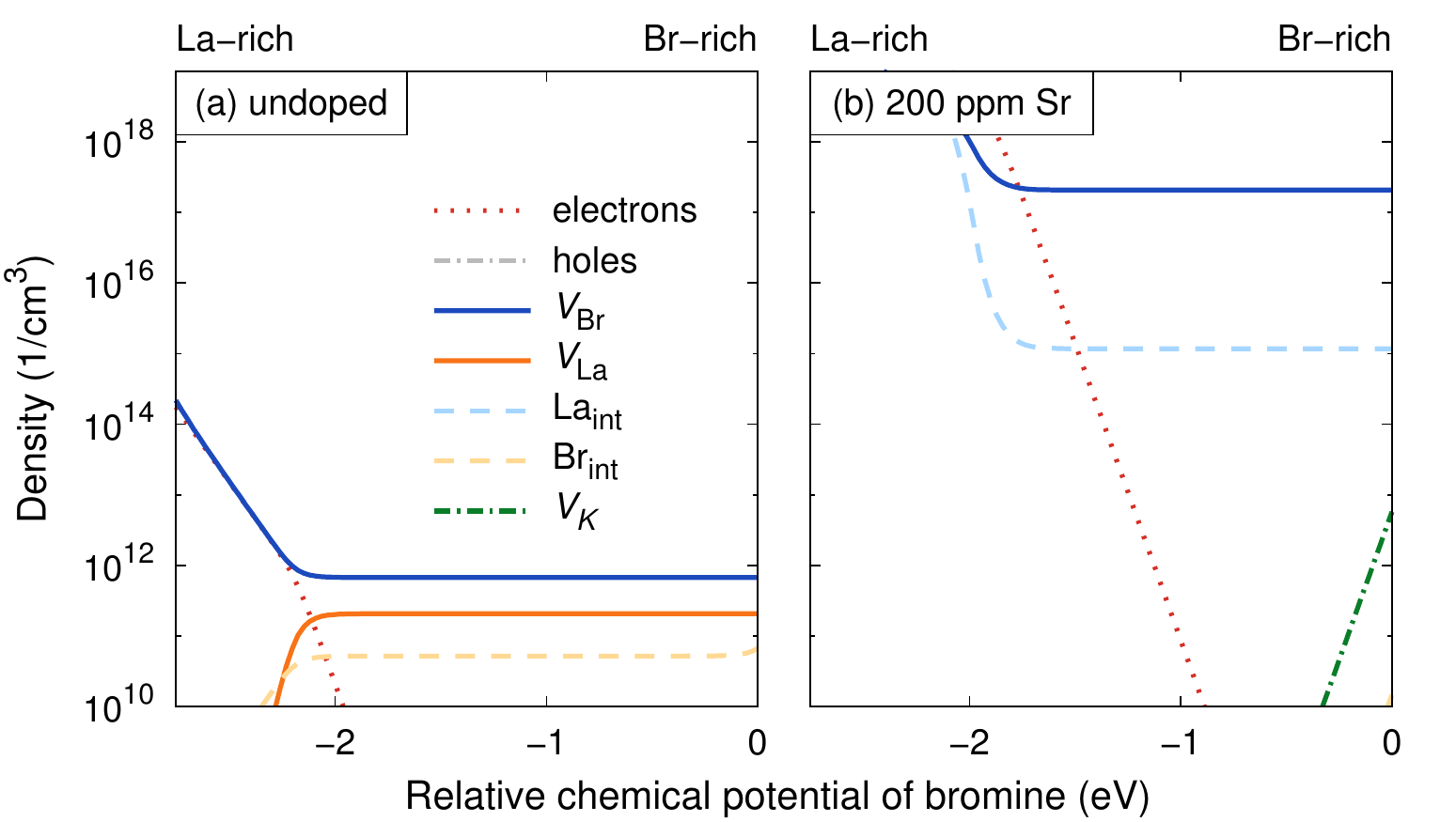}
\includegraphics[scale=\myscale]{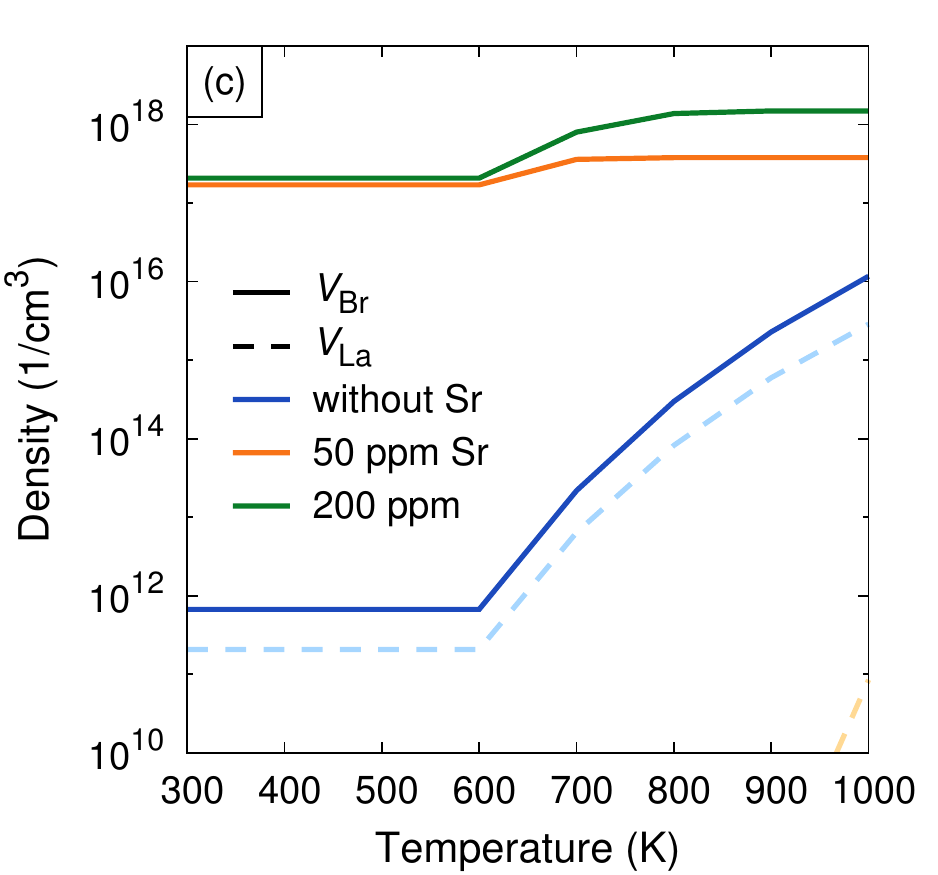}
  \caption{
    Defect and free charge carrier concentrations in (a) pure and (b) Sr-doped LaBr$_3$ as a function of chemical potential assuming full equilibrium at 600\,K (also compare Ref.~\onlinecite{AbeSadSch14}).
    (c) Concentrations of Br (solid lines) and La (dashed lines) vacancies as a function of temperature for different levels of Sr doping and assuming defect freezing at 600\,K.
    Note that incorporation of Sr leads to an increase in the Br vacancy concentration by several orders of magnitude.
  }
  \label{fig:defconc}
\end{figure*}

Using the pSIC method,\cite{SadErhAbe14} we have identified several self-trapped hole configurations (also known as $V_K$ centers). Analogously to the classic $V_K$ centers in NaI, the self-trapped polaron configurations in LaBr$_3$ involve a dimerization of two halide ions. We find that the Br--Br distance of the most energetically favorable $V_K$ center is 2.98\,\AA{} compared to 3.6\,\AA{} in the ideal lattice, see \fig{fig:structures}(d), with a binding/formation energy of $-0.3\,\eV$. A more detailed exposition on polaron binding energies and migration barriers in LaBr$_3$ will be published elsewhere.

Knowledge of the formation energies in combination with the charge neutrality condition, see \sect{sect:methodology}, allows one to compute defect concentrations as a function of the chemical boundary conditions as done previously in Ref.~\onlinecite{AbeSadSch14}. Here, it is exemplarily illustrated in \fig{fig:defconc}(a), which shows the dependence of defect concentrations and free charge densities on the relative chemical potential of Br, $\Delta\mu_\Br$, assuming full equilibrium at a temperature of 600\,K. In this representation,  $\Delta\mu_\Br\rightarrow 0$ corresponds to Br-rich conditions, implying that the material is equilibrated with respect to a Br-rich reservoir such as Br$_2$ gas, see Appendix~\ref{sect:eform}. In the La-rich limit, $\Delta\mu_\Br \rightarrow \frac{1}{3} H_f(\La\Br_3)$, which is equivalent to $\Delta\mu_\La\rightarrow 0$.

According to \fig{fig:defconc}(a) for the widest range of chemical conditions charge equilibrium is accomplished by La and Br vacancies, which act as acceptors and donors, respectively, or in Kr\"oger-Vink notation $[V_\La'''] \approx 3[V_\Br^\mydot]$. In the extreme La-rich limit the present calculations suggest that the La vacancies in this balance are replaced by free electrons, albeit at a small concentration. As the band gap is underestimated with respect to experiment (5.3 vs 5.9\,eV, compare \sect{sect:methodology}), the concentrations of free charge carriers are somewhat overestimated relative to defects. Being aware of this shortcoming we focus on the Br-rich limit from here on. The general conclusions drawn from our results are, however, entirely unaffected by this issue.

\subsection{Strontium}
\label{sect:strontium}

After the basic properties of intrinsic defects have been established, one can now explore the effect of Sr incorporation. The energetics of interstitial as well as substitutional defect configurations were considered as shown in \fig{fig:eform}(c,d), from which $\Sr_\La$ emerges as the dominant form. This defect acts as a singly charged acceptor $\Sr_\La'$ over the widest range of electron chemical potentials with an equilibrium transition level less than 0.2\,eV above the valence band maximum (VBM). Both the positioning of the transition level in the vicinity of the band edge and vanishingly small structural changes between charge states $q=-1$ and $0$ indicate that the defect is electronically shallow.

Under certain chemical conditions interstitial Br, which acts as a shallow donor, can also assume low formation energies. From a more detailed analysis of defect concentrations this defect is, however, found to occur generally in much smaller concentrations than $\Sr_\La$.

\begin{figure}
  \centering
\includegraphics[scale=\myscale]{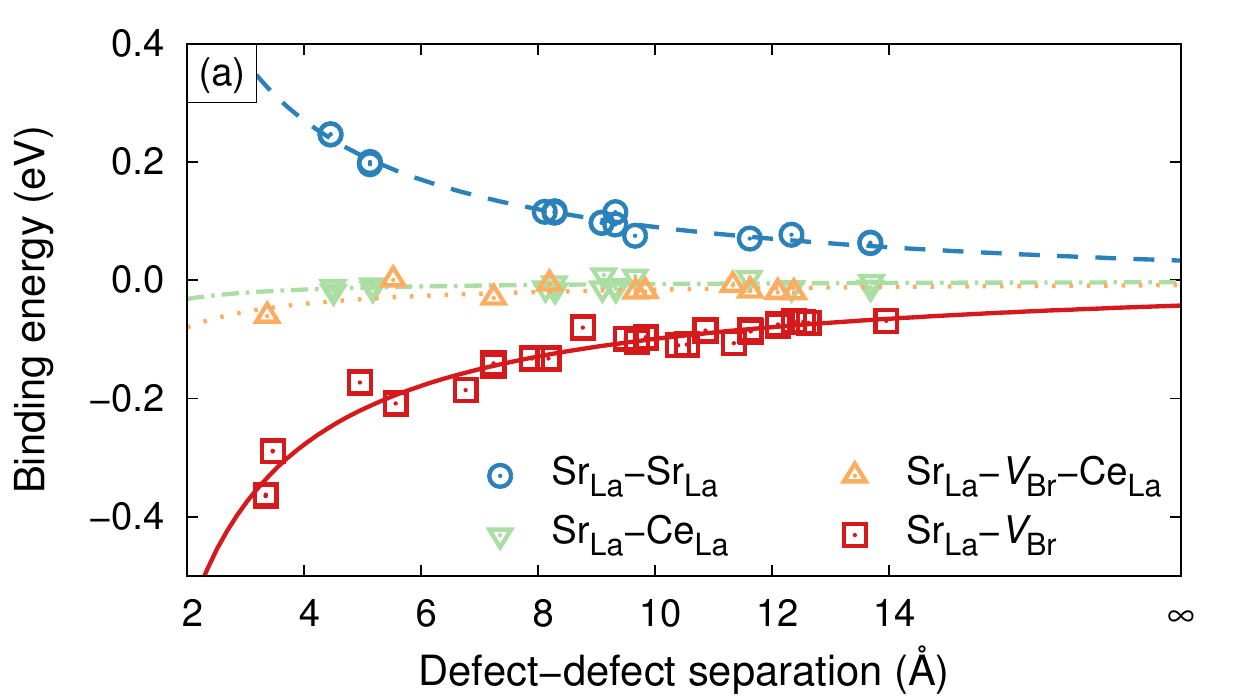}
\includegraphics[scale=\myscale]{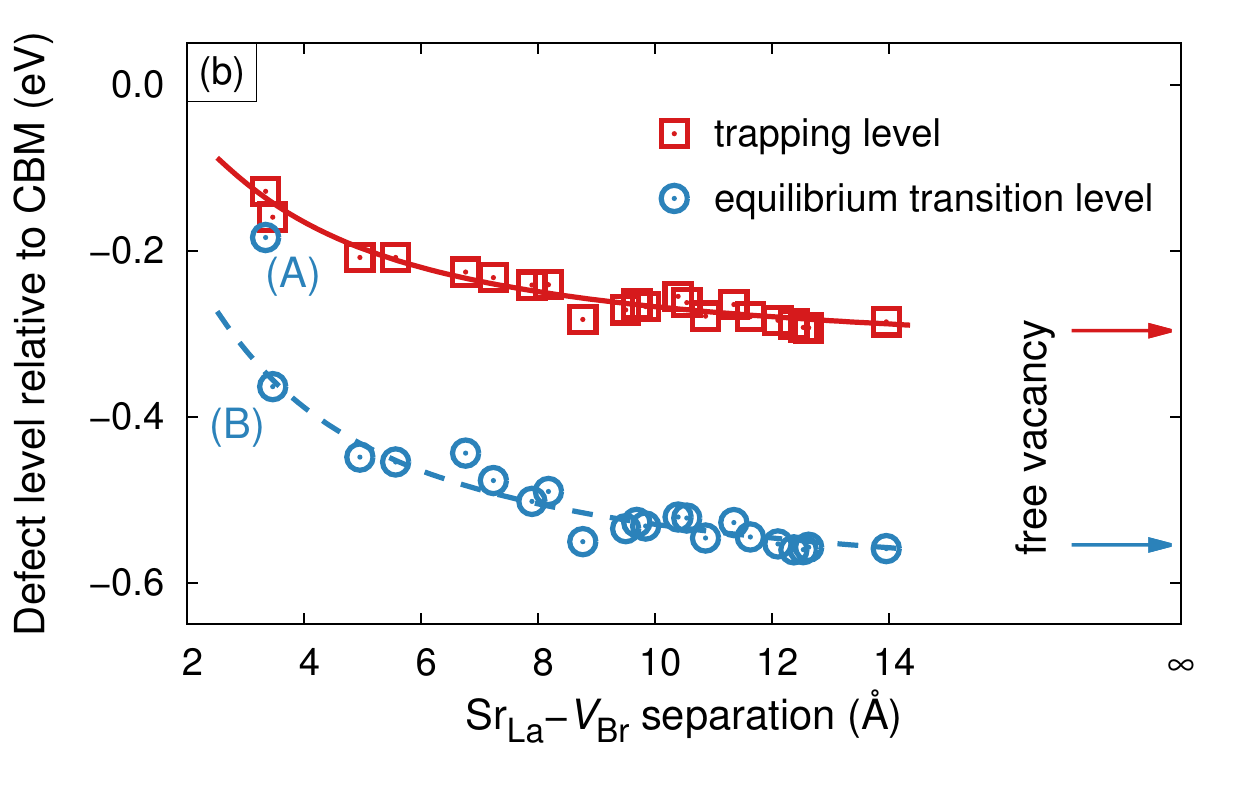}
\includegraphics[scale=\myscale]{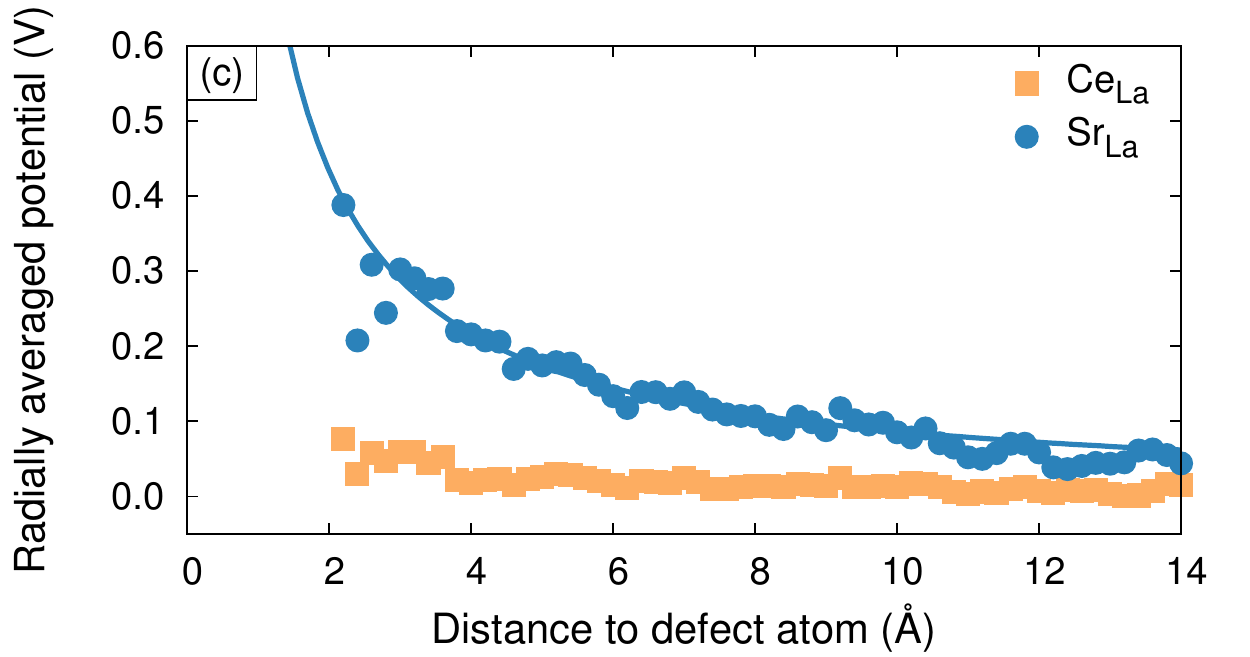}
  \caption{
    (a) Binding energies of Sr with various defects as a function of separation.
    The $\Sr_\La-V_\Br-\Ce_\La$ triple complex is composed of a nearest neighbor in-plane $\Sr_\La-V_\Br$ defect and $\Ce_\La$ at various distances.
    (b) Position of equilibrium transition and trapping levels of Br vacancy as a function of separation from $\Sr_\La$ (also compare Ref.~\onlinecite{AbeSadSch14}). The arrows on the right-hand side indicate the levels obtained for the free Br vacancy, compare \fig{fig:confcoord}. The data points marked (A) and (B) correspond to first nearest neighbor out-of-plane and in-plane configurations, respectively, compare \fig{fig:structures}.
    (c) Potential induced by $\Ce_\La^\times$ and $\Sr_\La'$ calculated by spherically averaging the difference in local potential between defect and ideal cell. The orange line is a fit to $1/r$.
  }
  \label{fig:ebind}
\end{figure}

The effect of Sr doping on intrinsic defect concentrations is exemplarily shown for a temperature of 600\,K in \fig{fig:defconc}(b). The vast majority of Sr is incorporated as $\Sr_\La$ and as an acceptor is balanced by Br vacancies, i.e. $[\Sr] \approx [\Sr_\La'] \approx [V_\Br^\mydot]$. As typical doping concentrations of Sr are between 50 and 200 ppm \cite{AleHaaKho13} this implies that co-doping with Sr leads to a substantial increase in the concentration of Br vacancy compared to pure LaBr$_3$.

Due to the limited mobility of atoms (and thus defects) at low temperatures crystalline materials are usually not in full defect equilibrium at low temperatures, say near room temperature. Rather defect concentrations are ``frozen in'' as the material is cooled down after manufacturing. A representative ``freezing'' temperature of 600\,K was assumed to generate \fig{fig:defconc}(c), which shows the evolution of the concentrations of intrinsic defects both for pure and doped materials. The data demonstrates that Sr doping can be expected to increase the Br vacancy concentration by up to five orders of magnitude compared to (nominally) pure material.

The opposite charge states of $V_\Br^\mydot$ and $\Sr_\La'$ cause mutual attraction as already shown in Ref.~\onlinecite{AbeSadSch14}. The interaction strength is quantified in \fig{fig:ebind}(a) revealing a binding energy of $-0.3\,\eV$ for the nearest neighbor $(\Sr_\La V_\Br)^\times$ complex.\footnote{We here adopt the convention that negative binding energies indicate attraction.} A closer inspection of the electronic structure of the complex reveals that the defect levels associated with the Br vacancy are shifted closer to the CBM by up to 0.4\,eV compared to the isolated vacancy, see \fig{fig:ebind}(b). The shift of the defect level can be rationalized by considering that $\Sr_\La'$ (unlike for example $\Ce_\La^\times$) introduces a point charge-like electrostatic potential that shifts the {\em local} energy scale, see \fig{fig:ebind}(c). Localized states such as the $V_\Br$ defect level are sensitive to this shift whereas the delocalized states that make up the valence and conduction bands are unaffected, causing an effective upward shift of the vacancy level.

It should be noted that there are two distinct first nearest neighbor $\Sr_\La-V_\Br$ configurations, which correspond to the Sr vacancy being oriented out-of-plane (point A) and in-plane (point B) relative to each other, respectively, compare \fig{fig:structures}. In spite of the very similar separation between vacancy and Sr, the out-of-plane complex has an equilibrium transition level that is 0.2\,eV closer to the CBM than in the in-plane geometry. This difference arises from the local constraints on relaxation that affect the two configurations differently. A similar effect is also observed in connection with the Stokes shifts associated with $\Ce_\La$-vacancy complexes, see \sect{sect:optical_signatures}.

In short, the incorporation of Sr in LaBr$_3$ thus (i) increases the Br vacancy concentration by several orders of magnitude and (ii) reduces the separation between vacancy level and CBM. These two effects have important implications with respect to understanding the improved scintillation response of Ce/Sr co-doped LaBr$_3$ as will be discussed in detail in \sect{sect:discussion}. Another important factor in this regard that should be mentioned here is the very small lattice distortion that occurs upon Sr incorporation, which will be revisited in \sect{sect:solubilities}.

\subsection{Cerium}

We now move on to consider the incorporation of Ce. Substitutional $\Ce_\La$ is energetically clearly preferred over Ce interstitial and $\Ce_\Br$ configurations. It remains neutral over the widest range of $\mu_e$ corresponding to $\Ce^{3+}$-$4f^15d^0$. If $\mu_e$ drops below $\epsilon_{tr}=0.37\,\eV$ the $4f$ level is depopulated leading to a configuration that corresponds to $\Ce^{4+}$-$4f^05d^0$. This range of electron chemical potentials is, however, typically not observed in experimental settings as demonstrated by the absence of $\Ce^{4+}$ signatures \cite{AleWebKra14}. Under La-rich conditions the formation energy difference between interstitial and substitutional Ce is reduced but $\Ce_\La$ remains the most dominant defect. The formation energy for $\Ce_\La^\times$ is $-0.71\,\eV$ assuming equilibrium with CeBr$_3$ regardless of chemical condition. The negative formation energy indicates full solubility of Ce in LaBr$_3$, which is compatible with the large amounts of Ce that are routinely substituted into the material.

As shown in \fig{fig:ebind}(a), the interaction of $\Ce_\La$ with other defects is very weak. Since $\Ce_\La$ is furthermore neutral under all relevant conditions, its effect on the charge neutrality condition and the concentrations of other defects is negligible.

\begin{table*}
  \centering
  \caption{
    Comparison of calculated and experimental data for Ce absorption and emission ($4f^15d^0 \leftrightarrow 4f^05d^1$). 
    Two values are given in the emission column corresponding to final states of $^2F_{7/2}$ and $^2F_{5/2}$, respectively.
    Experimental data from Ref.~\onlinecite{AleWebKra14}.
    IP: in-plane relative to $V_\Br$; OP: out-of-plane relative to $V_\Br$.
    $\Delta E$: total energy difference;
    $\Delta E_{rlx}^{exc}$: relaxation energy on excited state PES;
    $\Delta E_{rlx}^{gs}$: relaxation energy on ground state PES;
    $a_{\text{FC}}$: ionic relaxation along the reaction path.
  }
  \label{tab:optics}
  \begin{ruledtabular} \begin{tabular}{llcccccccc}
      \multicolumn{2}{l}{Site} & Figure & $\Delta E$ & Excitation &  Emission & Stokes shift & $\Delta E_{rlx}^{exc}$ & $\Delta E_{rlx}^{gs}$ & $a_{\text{FC}}$ \tabularnewline
      & & & & (eV) & (eV) & (eV) & (eV) & (eV) & (\AA) \tabularnewline
      \hline\\[-9pt]
      \multicolumn{4}{l}{Calculation} \tabularnewline
      & $\Ce_\La$                 &  & & 3.56 &  2.78 / 3.13 & 0.43 & 0.19 & 0.24 & 0.35 \tabularnewline
      & $\Sr_\La-V_\Br-\Ce_\La$ \tabularnewline
      & \phantom{x} $\Ce_\La$ IP, $\Sr_\La$ OP & \ref{fig:structures}(e) & 0.00 & 3.24 & 2.73 / 3.08 & 0.16 & 0.07 & 0.09 & 0.24 \tabularnewline
      & \phantom{x} $\Ce_\La$ OP, $\Sr_\La$ OP & \ref{fig:structures}(f) & 0.02 & 3.33 & 2.73 / 3.08 & 0.25 & 0.11 & 0.14 & 0.34 \tabularnewline
      & \phantom{x} $\Ce_\La$ OP, $\Sr_\La$ IP   & \ref{fig:structures}(g) & 0.09 & 3.28 & 2.55 / 2.90 & 0.38 & 0.15 & 0.23 & 0.49 \tabularnewline[6pt]
      \multicolumn{4}{l}{Experiment (Ref.~\onlinecite{AleWebKra14})} \tabularnewline
      & I             & &  & 4.03           & 3.47 / 3.19 & 0.56 \tabularnewline
      & II            & &  & 3.59 & 3.36 / 3.10 & 0.24 \tabularnewline
      & III           & &  & 3.47 & 3.27 / 3.10 & 0.21 \tabularnewline
  \end{tabular} \end{ruledtabular} 
\end{table*}
 
\begin{figure}
  \centering
\includegraphics[scale=\myscale]{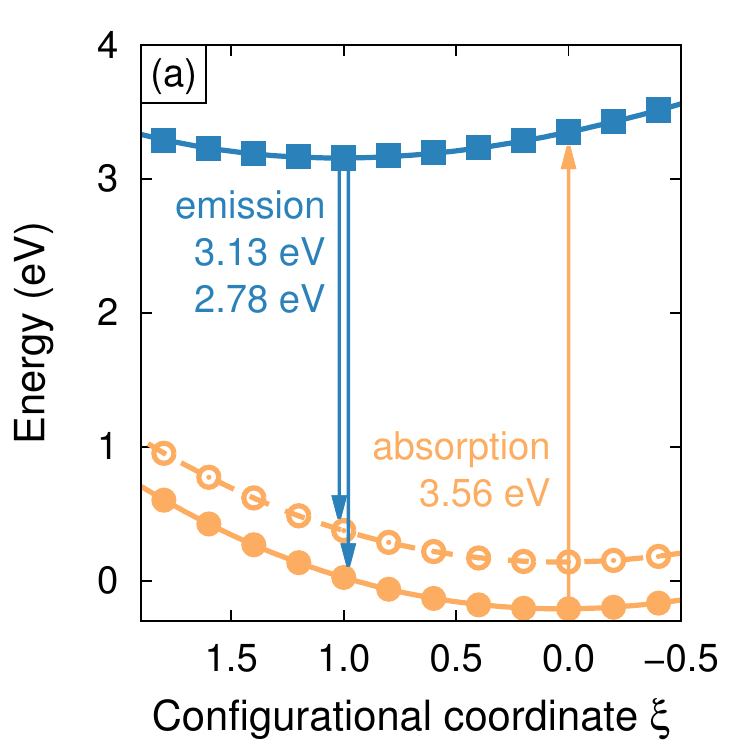}
\includegraphics[scale=0.3]{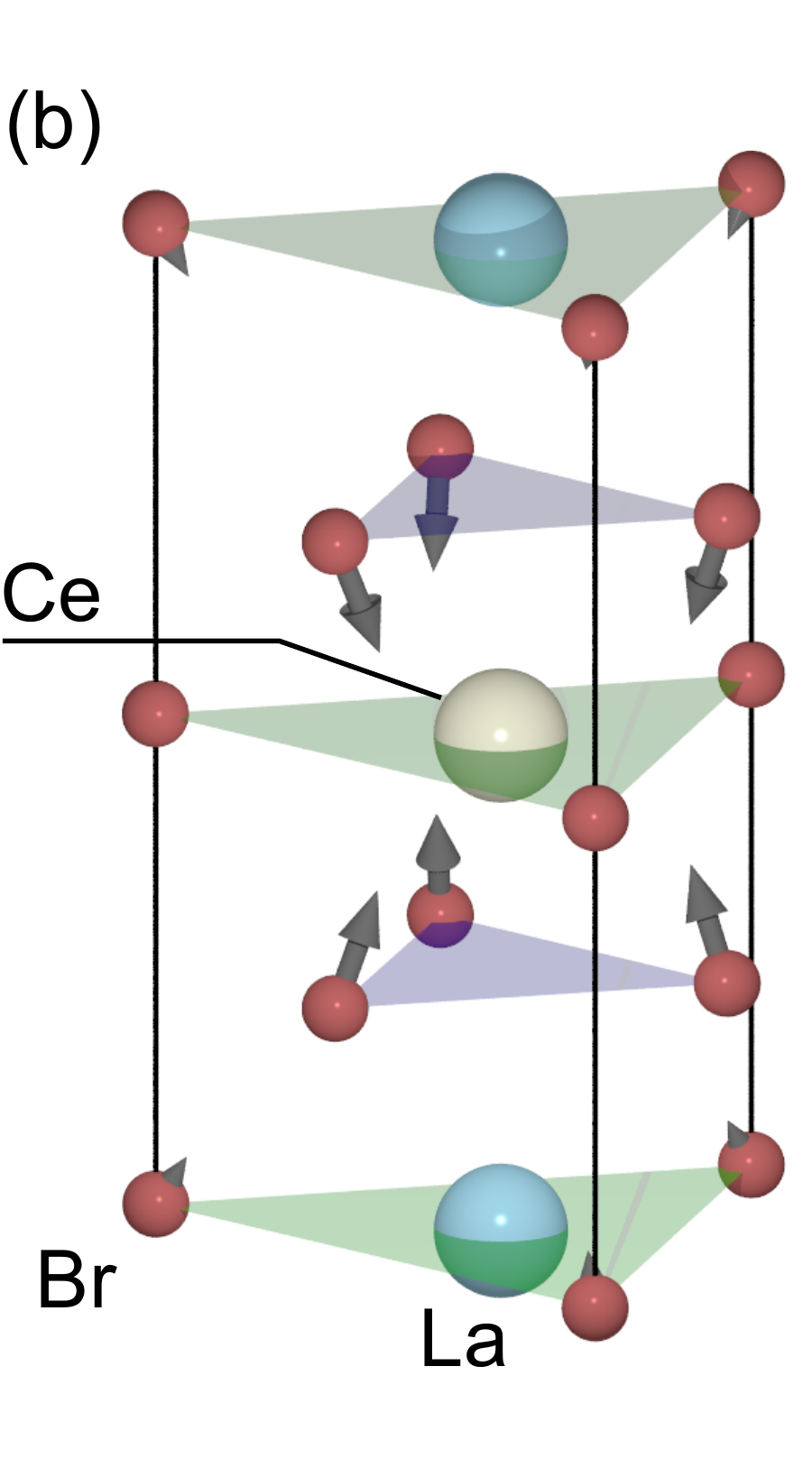}
  \caption{
    (a) Configuration coordinate diagram illustrating the relation between ground and excited state of $\Ce_\La$ in its neutral charge state. The ground and excited state PES correspond to the electronic configurations $\Ce^{3+}$-$4f^15d^0$ and $\Ce^{3+}$-$4f^05d^1$, respectively. The spin-orbit splitting of the ground state PES is represented by filled and open circles.
    (b) Schematic representation of the displacements associated with the configurational coordinate. The primary change pertains to the six out-of-plane Br neighbors of the Ce site, which relax inward by about 0.1\,\AA{} going from the ground to the excited state ionic configuration while the in-plane neighbor positions are almost unaffected.
  }
  \label{fig:confcoord_Ce}
\end{figure}

Upon high energy radiation or absorption above the band gap, Ce can be excited corresponding to the transition from $\Ce^{3+}$-$4f^15d^0$ to $\Ce^{3+}$-$4f^05d^1$. The latter is associated with the emergence of electronic levels close to the CBM. They are predominantly of Ce-$5d$ character and strongly hybridized with the neighboring La-$5d$ states. The calculated ground and excited state PES for $\Ce_\La$ are shown in the configuration coordinate diagram \fig{fig:confcoord_Ce}(a). While in the ground state configuration the ionic positions are almost unchanged compared to the perfect lattice, in the excited state the nearest neighbor Br atoms move inward by about 0.1,\AA.

We obtain excitation and emission energies of 3.56 and 3.13/2.78\,eV, where the latter two values correspond to final states $^2F_{5/2}$ and $^2F_{7/2}$, respectively, see \tab{tab:optics}. These data underestimate the experimental values of 4.03 and 3.47/3.19\,eV (site I in \tab{tab:optics}) by 0.47 and 0.34/0.41\,eV, respectively, which is expected given the well known band gap error of DFT. The error is, however, systematic and affects all transitions considered here in approximately the same way. The difference between the excitation and the larger emission energy gives the Stokes shift, for which the calculations yield 0.43\,eV in good agreement with the experimental value of 0.55\,eV. The relation between the ground and excited state landscapes is further illustrated in the configuration coordinate diagram \fig{fig:confcoord_Ce}.

The excitation and emission lines as well as the Stokes shift for Ce in LaBr$_3$ were also calculated by Andriessen and co-workers \cite{AndKolDor07}, who obtained a value of 0.42\,eV for the latter in close agreement with the present calculations in spite of differences in the computational approach regarding both pseudopotentials, the treatment of $4f$-states, and the description of the excited state. In contrast to those calculations, however, we do not obtain an asymmetric relaxation pattern for the excited state, in which the Ce ion is displaced along one of the Ce--Br bonds, described in Ref.~\onlinecite{AndKolDor07} and interpreted as a pseudo Jahn-Teller distortion. Rather we obtain a symmetric displacement pattern as described above even if the structural optimization is started from the asymmetric structure from Ref.~\onlinecite{AndKolDor07}. We conjecture that this discrepancy is related to the lack of DFT+$U$ correction terms in Ref.~\onlinecite{AndKolDor07}, which lead to both an erroneous ordering of La-$4f$ and $5d$ states \cite{AbeSadErh12} and partial occupancy of the Ce-$4f$ level in the $\Ce^{3+}$-$4f^15d^0$ configuration.

\subsection{Optical signatures of Ce complexes}
\label{sect:optical_signatures}

Introduction of Sr in the lattice is associated with the emergence of additional features in the optical spectra \cite{AleWebKra14}, indicating the existence of at least two additional Ce sites characterized by different Stokes shifts and also different absorption and emission wavelengths. As discussed above, Ce does not exhibit a propensity to form stable defect clusters. On the other hand, we have shown that $\Sr_\La$ and $V_\Br$ have a strong tendency to bind in two separate configurations. The Ce dopant level in detector material can be as high as 5\%. Even without defect-defect interactions, this implies that statistically there is a large probability of 14\%\ for a Ce to be in the immediate vicinity of a $\Sr_\La-V_\Br$ cluster. By inspection we find three different nearest neighbor $\Ce_\La-\Sr_\La-V_\Br$ triple clusters, depicted in \fig{fig:structures} (e--g). Each of these is characterized by the position [out-of-plane (OP) or in-plane (IP)] of the Br vacancy relative to Sr and Ce. 

For the $\Ce_\La$ IP, $\Sr_\La$ OP cluster, see \fig{fig:structures}(e), we obtain excitation and emission energies of 3.24 and 3.08/2.73\,eV. This corresponds to a Stokes shift of 0.16\,eV, much smaller than the predicted value for $\Ce_\La$ (see \tab{tab:optics}). Similarly, the Stokes shifts of $\Ce_\La$ OP, $\Sr_\La$ OP [\fig{fig:structures}(d)] and $\Ce_\La$ OP, $\Sr_\La$ IP [\fig{fig:structures}(e)] are 0.25 and 0.38\,eV, respectively. In the nomenclature of Ref.~\onlinecite{AleWebKra14} we thus tentatively assign the first cluster to site III, and the remaining two to site II. 

The smaller Stokes shifts for the triple clusters can be rationalized in terms of smaller structural relaxations in the excited $4f^05d^1$ state as measured by $a_{\text{FC}}$ of \eq{eq:confcoord}, see \tab{tab:optics}. We indeed find a direct correlation between the size of the Stokes shift and the amount of relaxation. In effect, this confirms the hypothesis of Refs.~\onlinecite{DorAleKho13} that the Ce excitation has a weaker influence on the geometrical structure if another defect is in its vicinity.

\subsection{Solubility analysis}
\label{sect:solubilities}

The detailed investigation of Sr related defects in Sects.~\ref{sect:strontium} and \ref{sect:optical_signatures} was motivated by the improvement of energy resolution observed for LaBr$_3$:Ce co-doped with Sr. Similar effects were observed with Ca and Ba whereas doping with Li, Na, or Mg does not improve the scintillation response and can even be detrimental \cite{AleBinKra13}. To resolve these observations, the investigation of dopant related defects was extended to cover both the alkaline (Li, Na, K, Rb, Cs) and alkaline earth (Be, Mg, Ca, Sr, Ba) groups. It comprised the same configurations and charge states that were already described in \sect{sect:strontium}. Defect formation energies were computed with respect to the respective bromide compounds (see Appendix~\ref{sect:chemical_bounds}, Br-rich conditions), which are commonly used to introduce the dopants in the synthesis \cite{AleBinKra13}.

\begin{figure}
  \centering
\includegraphics[scale=\myscale]{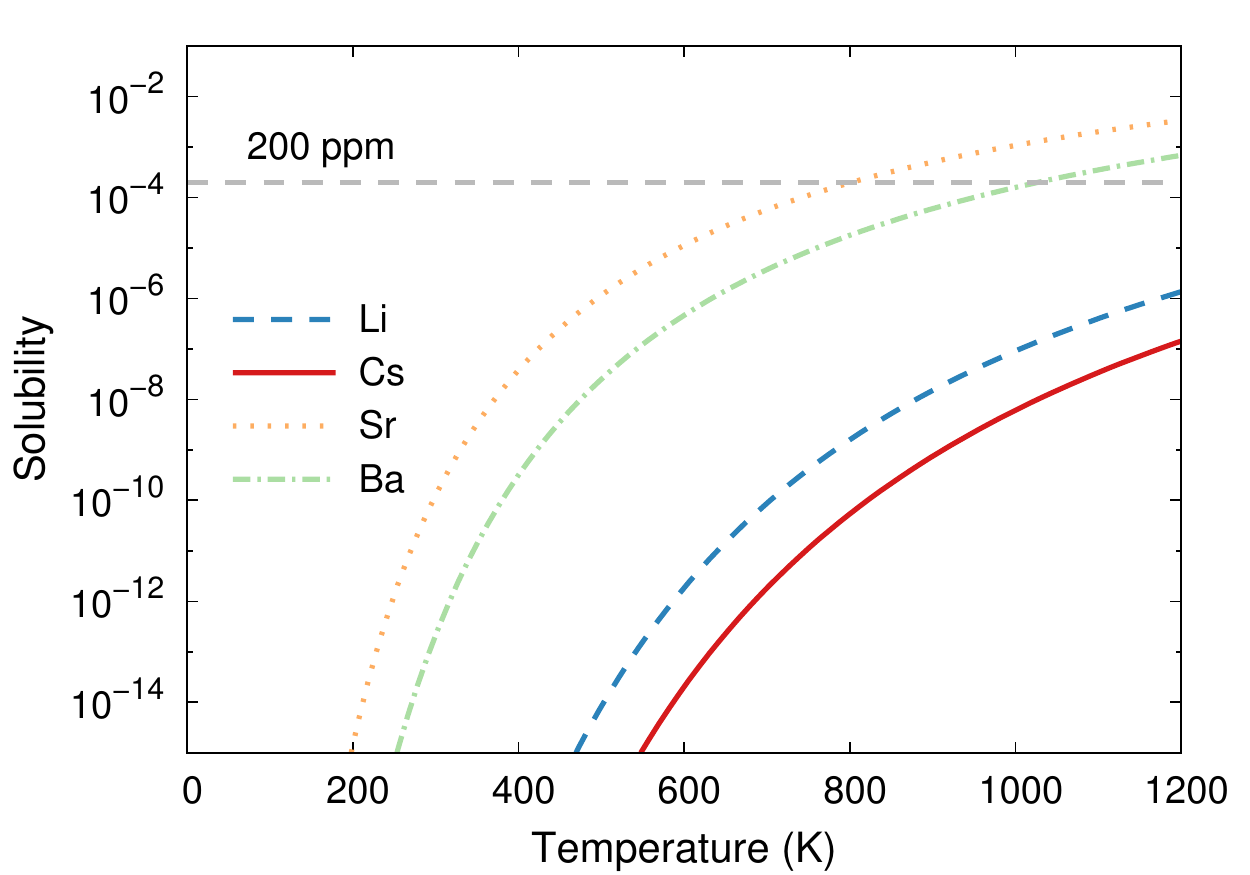}
  \caption{
    Solubility of selected dopants computed using \eq{eq:defconc} and the formation energies for substitution on La.
    Li and Cs demark the upper and lower limit of solubilities for alkaline doping.
  }
  \label{fig:solubility}
\end{figure}

The results with regard to the dependence of the electron chemical potential are qualitatively similar to \fig{fig:eform}(c,d) for all dopants considered here. The formation energies of different dopants shift, however, relative to each other. Under Br-rich conditions the Fermi level is located in the lower part of the band gap. Under these circumstances, substitution on La is energetically the most favorable form for Ca, Sr, and Ba, whereas interstitial configurations are preferred for the Be and Mg. For the alkaline metals, substitution on La sites is preferred but the formation energies are noticeably larger than for the alkaline earth metals. The solubility can be obtained from the formation energies for substitution on La and \eq{eq:defconc}, which yields the data shown in \fig{fig:solubility}. It is apparent that the solubilities of alkaline metals are generally several orders of magnitude smaller than for the alkaline earth metals.

Experimentally, only Ba, Sr, and Ca are found to achieve an improvement in scintillator response. This behavior can be understood on the basis of these results as follows:
While the alkaline metals appear to be electronically and geometrically well matched to the LaBr$_3$ lattice (small distortions, shallow levels), their solubilities are too small to accomplish thermodynamically significant incorporation. Be and Mg exhibit small formation energies and thus relatively higher solubilities. They are, however, associated with substantial lattice distortions and interstitial-like configurations, which gives rise to carrier scattering and an overall degradation of electronic conduction. In the end, only the heavy alkaline earth metals (Ca, Sr, Ba) combine excellent electronic and geometric match (small strains, shallow levels) with sufficiently large solubilities, providing a rationale for the success of these dopants.

\section{Discussion and conclusions}
\label{sect:discussion}

To summarize, a comprehensive investigation of intrinsic and extrinsic defects in LaBr$_3$ has been conducted on the basis of first principles calculations within density functional theory. The vacancies, $V_\Br$ and $V_\La$, were found to be the most dominant intrinsic donor and acceptor defects, respectively. The equilibrium concentration of $V_\Br$ in the nominally pure (undoped) case is about $10^{12}\,\cm^{-3}$, assuming a freezing-in temperature of 600\,K. The Br vacancy is associated with deep equilibrium transition, trapping, and detrapping levels located several tenths of an eV below the CBM.

Sr substitutes for La and acts as a shallow acceptor. Assuming a Sr dopant concentration of 200\,ppm, charge compensation to ensure overall neutrality increases the $V_\Br$ concentration by up to five orders of magnitude. The coulombic attraction between these two species causes formation of stable $\Sr_\La-V_\Br$ complexes with a binding energy of about $-0.3\,\eV$. Upon this reaction the defect level associated with the vacancy is shifted by up to 0.4\,eV toward the CBM. Thus, incorporation of Sr introduces a multitude of shallow electron traps.

Cerium substitutes for La in a local 3+ charge state and the calculations support the experimental conclusion that no appreciable amount of Ce 4+ is present \cite{AleWebKra14}. Since Ce substitutes another triply ionized atom it has no electrostatic interactions with other defects and hence does not prefer to bind to either $\Sr_\La$, $V_\Br$, or their complexes. The high Ce concentrations used in practice, however, imply that there is very high probability of 14\%\ for a Ce atom to be in the first neighbor shell of a $\Sr_\La-V_\Br$  complex. This picture is supported by the good agreement between experimentally observed and computed Stokes shifts for the $4f \leftrightarrow 5d$ transition of isolated as well as complexed Ce.

In the Ce $4f^15d^0$ configuration, the empty $5d$ levels reside inside the conduction band while the occupied $4f$ state is associated with a deep level inside the gap. Conversely, in the excited $4f^05d^1$ configuration, an occupied $5d$ level is present inside the band gap. Excluding the possibility of energy transfer from an exciton, we note that the Ce is most likely to be excited via a sequential hole and electron capture. In the Ce-only doped case, which is almost identical to the nominally pure material, no substantial amount of electron traps can be expected. Furthermore, since the Ce $4f$ level is very deep, it is natural to assume that the initial hole capture is the rate limiting step and thus that there is no fast mechanism to reduce the electron/hole density. This certainly favors Auger recombination, the rate of which has a cubic dependence on the excitation density and which has been shown to be a major quenching channel at the relevant time scales for halide scintillators in $z$-scan experiments \cite{WilGriLi13,GriUceBur13}.

Conversely, by co-doping with Sr, the electron trap levels not only become more shallow, which allows for faster trapping/detrapping rates, but the trap density increases by several orders of magnitude in the form of $\Sr_\La-V_\Br$ complexes. If these traps are active during the initial thermalization stage (2--10\,ps in halide systems\cite{GriUceBur13}) they will effectively reduce the free electron density. As a result, a larger number of holes will remain available for ionization of cerium activators as the probability for quenching of electron-hole pairs via the Auger mechanism should be greatly reduced. Recent picosecond optical absorption experiments have shown that energy transfer to europium activators in SrI$_2$:Eu may be as fast as 400\,fs \cite{UceBizBur14}, which demonstrates that very fast capture is indeed possible. As each defect complex will be in close proximity to a Ce atom, once any of the nearby Ce atoms captures a hole, Coulombic attraction serves as a driving force for transferring the electron from the complex to the activator. This suggests that non-linear quenching is reduced at the cost of longer decay-times. In fact, two of the three cerium sites discussed in Ref.~\onlinecite{DorAleKho13} are associated with very long decay times ranging from 60 to 2500\,ns while accounting for 20-45\,\%\ of the total light output \cite{AleWebKra14}.

The mechanism outlined above demonstrates that co-doping of wide gap materials, in particular scintillators, provides an efficient means for managing charge carrier populations under out-of-equilibrium conditions. In the case of LaBr$_3$:Ce,Sr the co-dopant manipulates not only the concentrations but also the electronic properties of {\em intrinsic} defects (specifically $V_\Br$) without introducing additional gap levels. This leads to the presence of shallow electron traps that can localize charge carriers on a nanosecond time scale, effectively deactivating charge carrier recombination channels. The principles of this mechanism are therefore not specific to the material considered here but can be adapted for controlling charge carrier recombination in other wide gap materials.

\acknowledgments

We acknowledge fruitful discussions with S. Payne, G. Bizarri, and R. T. Williams.  This work was performed under the auspices of the U.S. Department of Energy by Lawrence Livermore National Laboratory under Contract DE-AC52-07NA27344 with support from the National Nuclear Security Administration Office of Nonproliferation Research and Development (NA-22). Specifically, modeling of excited Ce states was supported by the Laboratory Directed Research and Development Program, Project. No. 13-ERD-038, at Lawrence Livermore National Laboratory. P.E. acknowledges funding from the Knut and Alice Wallenberg Foundation and the European Research Council in the form of a Marie Curie Career Integration Grant. Com\-puter time allocations by the Swedish National Infrastructure for Computing at NSC (Link\"oping) and C3SE (Gothenburg) are gratefully acknowledged.

\appendix

\section{Defect thermodynamics}
\label{sect:eform}

In the dilute limit the equilibrium concentration of a defect depends on its free energy of formation, $\Delta G_f$, via
\begin{align}
   c = c_0 \exp\left(-\Delta G_f / k_B T \right),
   \label{eq:defconc}
\end{align}
where $c_0$ is the concentration of possible defect sites. The formation free energy $\Delta G_f$ is usually approximated by the formation energy $\Delta E_f$, which is legitimate if the vibrational entropy and the pressure-volume term are small \cite{AbeErhWil08}. The formation energy $\Delta E_f$ of a defect in charge state $q$ is given by \cite{QiaMarCha88, ZhaNor91, ErhAlb07}
\begin{align}
  \Delta E_f &= E_\text{def} - E_\text{id} + q (\epsilon_{\VBM} + \mu_e) - \sum_i \Delta n_i \mu_i
  \label{eq:eform_raw}
\end{align}
where $E_\text{def}$ is the total energy of the system containing the defect and $E_\text{id}$ is the total energy of the ideal host. The second term describes the dependence on the electron chemical potential, $\mu_e$, which is measured with respect to the valence band maximum (VBM), $\epsilon_{\VBM}$. The formation energy also depends on the chemical potentials of the constituents as given by the last term, where the difference between the number of atoms of type $i$ in the ideal cell with respect to the defect cell is denoted by $\Delta n_i$. The chemical potential $\mu_i$ of constituent $i$ can be rewritten as $\mu_i = \mu_i^{bulk} + \Delta\mu_i$ where $\mu_i^{bulk}$ denotes the chemical potential of the standard reference state. Neglecting entropic contributions the chemical potentials of the reference phases can be replaced by their cohesive energies at zero Kelvin. Note that the summation in the last term also includes dopant or impurity species, whence one has to consider the source of the dopant or impurity atom when discussing formation energies and solubilities (see \sect{sect:solubilities}). The chemical potentials of La and Br are coupled to each other via the formation enthalpy of the compound, i.e., $\Delta\mu_\La + 3\Delta\mu_\Br = \Delta H_f$. This implies that specifying either $\Delta\mu_\La$ or $\Delta\mu_\Br$ is sufficient to determine the chemical equilibrium with respect to the host. Following common practice we refer to La and Br-rich conditions, which correspond to $\Delta\mu_\La=0\,\eV$ and $\Delta\mu_\Br=0\,\eV$, respectively.

\section{Thermodynamic boundary conditions}
\label{sect:chemical_bounds}

When calculating defect formation energies of intrinsic defects according to \eq{eq:eform_raw} it is sufficient to consider the chemical potentials for La and Br only as indicated above; specifically, for La-rich conditions
\begin{align}
  \Delta\mu_\La = 0 &\quad\text{and}\quad \Delta\mu_\Br = H_f(\La\Br_3)/3
  \intertext{whereas for Br-rich conditions}
  \Delta\mu_\Br = 0 &\quad\text{and}\quad \Delta\mu_\La=H_f(\La\Br_3).
\end{align}
Once extrinsic elements have to be taken into account additional conditions must be invoked. To be specific consider the case of Sr, which introduces one additional chemical potential, $\mu_\Sr = \mu_\Sr^0 + \Delta\mu_\Sr$, in \eq{eq:eform_raw}. One could assume the dopant to be in equilibrium with its elemental (metallic) form, which implies
\begin{align}
  \Delta\mu_\Sr = 0.
\end{align}
Sr and Br can, however, react to form SrBr$_2$ and it is therefore more appropriate to consider the equilibrium with respect to the compound, which is also used experimentally for introducing the dopant during synthesis \cite{AleBinKra13},
\begin{align}
  \Delta\mu_\Sr + 2\Delta\mu_\Br = H_f(\Sr\Br_2).
\end{align}
Since this reaction involves an element of the host the intrinsic boundary conditions (La and Br-rich) explicitly affect the condition for the chemical potential of Sr. In Br-rich conditions
\begin{align}
  \Delta\mu_\Br &= 0 \nonumber \\
  &\rightarrow \Delta\mu_\Sr = H_f(\Sr\Br_2)
\end{align}
while in the La-rich limit
\begin{align}
  \Delta\mu_\Br &= \frac{1}{3} H_f(\La\Br_3) \nonumber \\
  &\rightarrow \Delta\mu_\Sr = H_f(\Sr\Br_2) - \frac{2}{3} H_f(\La\Br_3).
\end{align}
The extension to other elements is straightforward. For the chemical potentials of the alkaline metals for example one obtains
\begin{align}
  \Delta\mu_\Na &= H_f(\Na\Br)                            && \text{Br-rich} \\
  \Delta\mu_\Na &= H_f(\Na\Br) - \frac{1}{3} H_f(\La\Br_3) && \text{La-rich}.
\end{align}
These conditions are used in \sect{sect:solubilities} to determine the solubilities of various dopants in LaBr$_3$.

\section{Finite-size scaling}
\label{sect:scaling}

Given the small dielectric constant of LaBr$_3$ and the large defect charge states that need to be considered it is crucial to properly correct for both electrostatic and elastic image charge interactions. Various correction schemes have been proposed for this purpose but ambiguities remain \cite{KomRanPas12}. In the present work we therefore resort to finite-size scaling, which if computationally affordable is expected to yield the most reliable results. Finite-size scaling is most commonly based on ``simple'' multiples of the underlying primitive cell. For example for diamond and zincblende structures supercells based on simple cubic, body-centered cubic, and face-centered cubic unit cells are often used \cite{CasMir04, LanZun08, AbeErhWil08}. Since each of these cells is associated with a different Madelung constant the size dependence of for example the monopole-monopole correction, which is the leading electrostatic interaction term \cite{MakPay95}, will differ between these cells. It is therefore advantageous to consider scaling among a set of self-similar cells.

The direct application of this principle to the hexagonal unit cell of LaBr$_3$ would allow the construction of only a very small number of supercells, which in turn would limit the reliability of the finite-size scaling procedure. To resolve this situation we devised a systematic way to construct ``optimal'' supercells. Optimality here implies that we seek to find supercells that for a given size (number of atoms) most closely approximate a cubic cell shape. This approach ensures that the defect separation is large and that the electrostatic interactions exhibit a systematic scaling. (Recall that for example the monopole-monopole interaction is given by the Madelung constant, which is only dependent on the shape of the unit cell).

The cubic cell metric for a given volume $\Omega$ is
\begin{align}
  \mat{h}_\text{cub} &= \Omega^{1/3} \mat{I},
\end{align}
which in general does not satisfy the crystallographic boundary conditions. The $l_2$-norm provides a convenient measure of the deviation of any other cell metric from a cubic shape (``acubicity'')
\begin{align}
  \Delta_c(\mat{h}) &= ||\mat{h} - \mat{h}_\text{cub}||_2.
  \label{eq:acubicity}
\end{align}
Cell metrics that are compatible with the crystal symmetry can be written as integer multiples of the underlying primitive unit cell $\mat{h_p}$, i.e.
\begin{align}
  \mat{h} &= \mat{P} \mat{h_{p}} \quad \text{where} \quad \mat{P} \in \mathbb{Z}^{3\times 3}.
\end{align}
The optimal cell shape multiplier for a given cell size is then obtained as follows
\begin{align}
  \mat{P}_{opt} &= \arg\min_P\left\{\Delta_c(\mat{P}\mat{h_{p}}) | \det \mat{P} = N_{uc}\right\},
  \label{eq:Popt}
\end{align}
where $N_{uc}$ is the desired system size in multiples of the primitive unit. This approach is generally applicable and can be readily generalized to optimize toward other supercell shapes, e.g., face-centered or body-centered cubic.

\begin{figure}
\includegraphics[scale=\myscale]{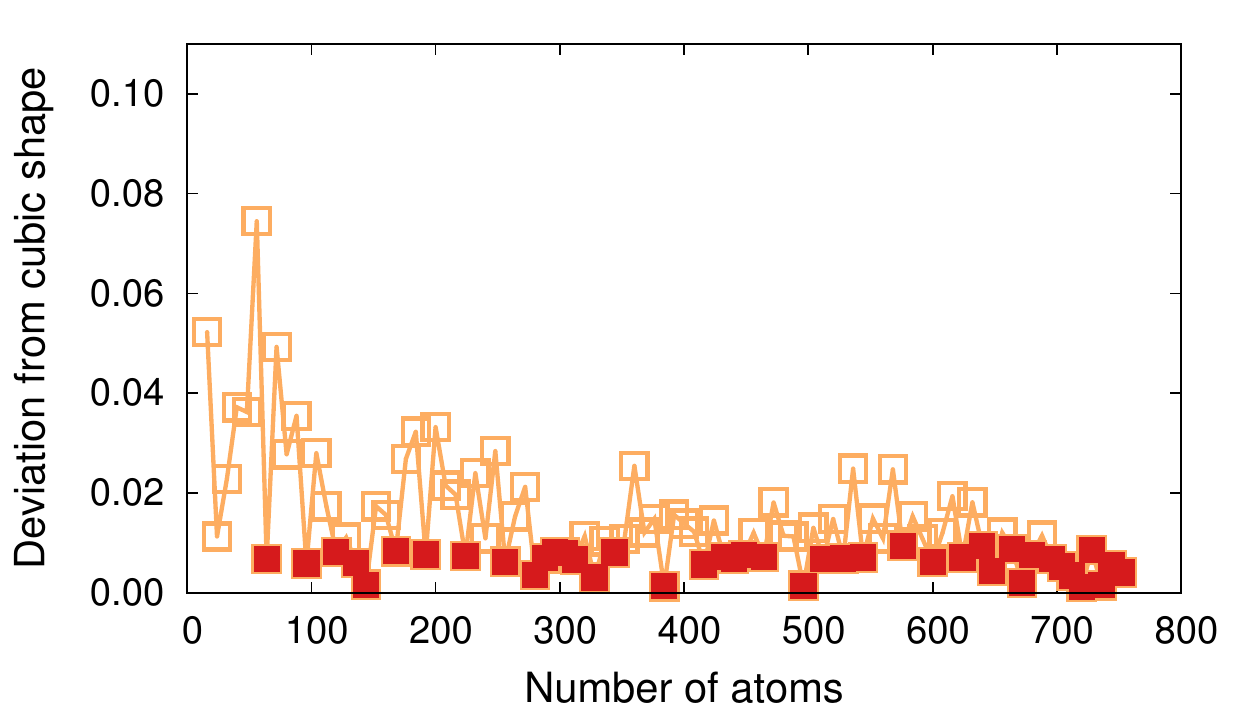}
  \caption{
    Lowest deviations from a cubic shape obtained for different cell sizes via \eq{eq:Popt} and \eq{eq:acubicity}.
  }
  \label{fig:supercells}
\end{figure}
A series of supercells was generated based on \eq{eq:Popt} for all possible sizes up to 752 atoms ($N_{uc}=94$). The lowest values of $\Delta_c$ achieved in this fashion are shown in \fig{fig:supercells}. We empirically find that supercells, for which $\Delta_c$ is lower than 0.02, are sufficiently close to a cubic shape for our purposes. The sizes for which this limit is reached are indicated by the filled red symbols in \fig{fig:supercells}.

Using supercells containing between 24 and 496 atoms defect calculations were carried out for Br and vacancies and antisites in charge states that were identified as relevant ones on the basis of earlier 96-atom cell calculations. The thus obtained configurations were analyzed as follows. Formation energies were computed using the thermodynamic formalism described in Appendix~\ref{sect:eform}. The formation energies obtained according to \eq{eq:eform_raw} are referred as ``raw'' data in the following. In addition we considered the effect of potential alignment (PA) and image charge corrections. The potential alignment correction amounts to a term $q v_\text{PA}$. In the present work we determined $v_\text{PA}$ by taking the difference of the electrostatic potential between defect and ideal supercell, with the potential being measured by a test charge at the ionic site farthest from the defect center. For the image charge correction we adopted the simplified correction described in Ref.~\onlinecite{LanZun08}, which involves the addition of a term $q^2 \frac{2}{3} |E_{mp}|$ where $E_{mp}$ is the electrostatic energy associated with a periodic array of point charges according to the supercell metric taking into account dielectric screening due to both electrons and ions. This correction term should reflect both monopole-monopole and monopole-quadrupole terms. The resulting expression for the formation energy is
\begin{align}
  \Delta \widetilde{E_f} &= \Delta E_f + q v_\text{PA} + q^2 \frac{2}{3} |E_{mp}|.
  \label{eq:eform_corr}
\end{align}
The leading terms in \eq{eq:eform_corr} should scale with $V^{1/3}$ and $V$ (or $N^{1/3}$ and $N$), therefore we also fit formation energies obtained from \eq{eq:eform_raw} to the following expression
\begin{align}
  \Delta E_f^\infty &= \Delta E_f(N) + a N^{-1/3} + b N,
  \label{eq:eform_fit}
\end{align}
where $\Delta E_f^\infty$, $a$ and $b$ were treated as fit parameters.

\begin{figure}
  \centering
\includegraphics[scale=\myscale]{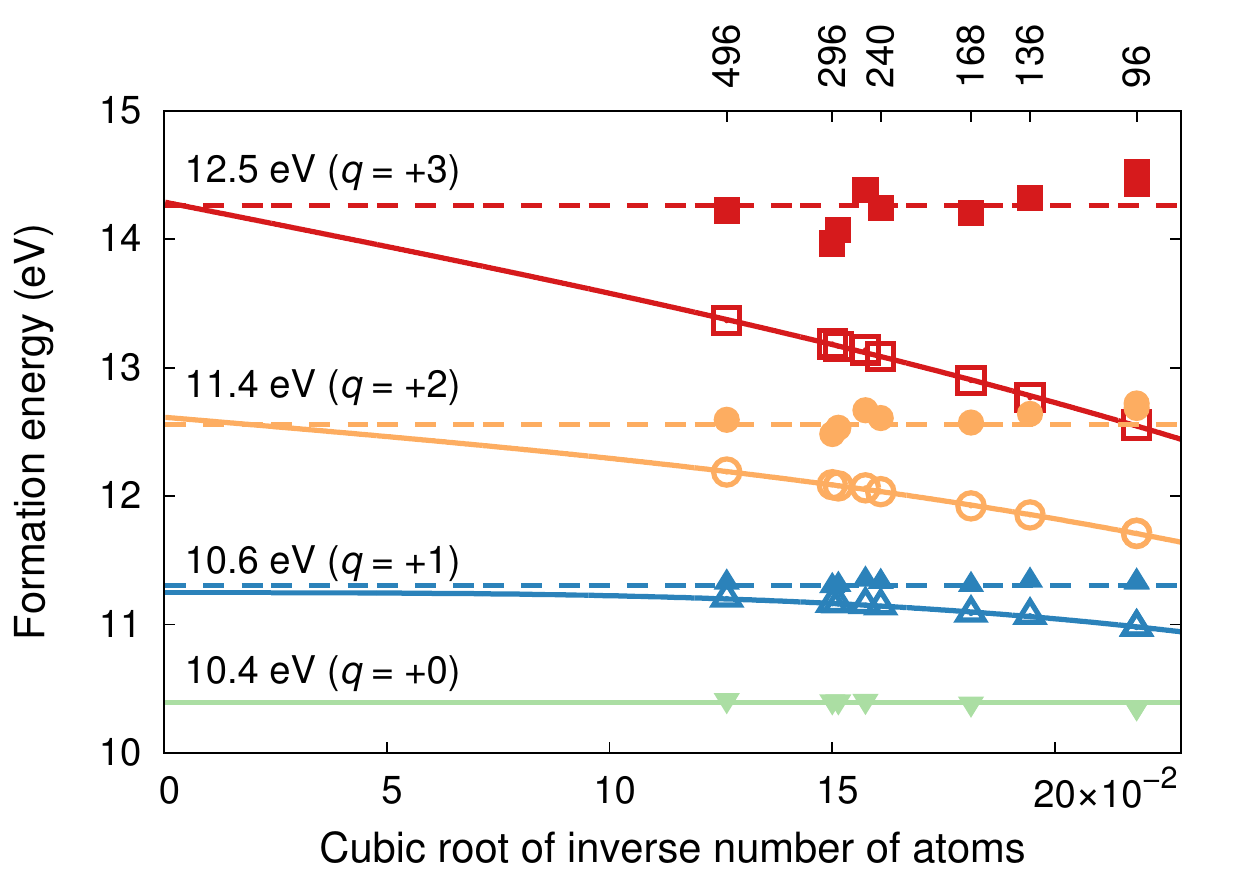}
  \caption{
    Finite-size scaling of formation energies of La vacancies in different charge states.
    Open symbols refer to formation energies calculated via \eq{eq:eform_raw} whereas filled symbols belong to formation energies subjected to potential alignment and image charge corrections according to \eq{eq:eform_corr}. Solid lines are fits to \eq{eq:eform_fit} and dashed lines indicate the average over the corrected formation energies. The numbers next to the lines indicate the formation energies obtained by extrapolation of the uncorrected data.
    In the case of the neutral vacancy ($q=0$) there are no potential alignment and image charge corrections.
    Formation energies have been computed for Br-rich conditions and an electron chemical potential $\mu_e=0.6\,\eV$, which was chosen to obtain a visual separation of the different charge states.
  }
  \label{fig:scaling_VLa}
\end{figure}

The results of this analysis are exemplified in \fig{fig:scaling_VLa}, which illustrates the scaling for La vacancies in various charge states. Analysis of e.g., Br vacancies, antisites, and interstitials yield very similar plots. In all cases we find that the combination of potential alignment and image charge corrections yields formation energies that approximate the infinite limit rather well. The accuracy of the thus corrected data is, however, limited by the accuracy associated with the determination of the potential alignment correction and the approximitive nature of the image charge correction. Note that the corrections do not account for elastic image interactions. The comparison of extrapolated values, which do include elastic effects, and corrected data, which do not, indicates that elastic interactions are, however, negligible in the present case. This conclusion is further supported by the fact that the formation energies of neutral defects are almost independent of system size for $N\geq 96$.


\begin{thebibliography}{48}%
\makeatletter
\providecommand \@ifxundefined [1]{%
 \@ifx{#1\undefined}
}%
\providecommand \@ifnum [1]{%
 \ifnum #1\expandafter \@firstoftwo
 \else \expandafter \@secondoftwo
 \fi
}%
\providecommand \@ifx [1]{%
 \ifx #1\expandafter \@firstoftwo
 \else \expandafter \@secondoftwo
 \fi
}%
\providecommand \natexlab [1]{#1}%
\providecommand \enquote  [1]{``#1''}%
\providecommand \bibnamefont  [1]{#1}%
\providecommand \bibfnamefont [1]{#1}%
\providecommand \citenamefont [1]{#1}%
\providecommand \href@noop [0]{\@secondoftwo}%
\providecommand \href [0]{\begingroup \@sanitize@url \@href}%
\providecommand \@href[1]{\@@startlink{#1}\@@href}%
\providecommand \@@href[1]{\endgroup#1\@@endlink}%
\providecommand \@sanitize@url [0]{\catcode `\\12\catcode `\$12\catcode
  `\&12\catcode `\#12\catcode `\^12\catcode `\_12\catcode `\%12\relax}%
\providecommand \@@startlink[1]{}%
\providecommand \@@endlink[0]{}%
\providecommand \url  [0]{\begingroup\@sanitize@url \@url }%
\providecommand \@url [1]{\endgroup\@href {#1}{\urlprefix }}%
\providecommand \urlprefix  [0]{URL }%
\providecommand \Eprint [0]{\href }%
\providecommand \doibase [0]{http://dx.doi.org/}%
\providecommand \selectlanguage [0]{\@gobble}%
\providecommand \bibinfo  [0]{\@secondoftwo}%
\providecommand \bibfield  [0]{\@secondoftwo}%
\providecommand \translation [1]{[#1]}%
\providecommand \BibitemOpen [0]{}%
\providecommand \bibitemStop [0]{}%
\providecommand \bibitemNoStop [0]{.\EOS\space}%
\providecommand \EOS [0]{\spacefactor3000\relax}%
\providecommand \BibitemShut  [1]{\csname bibitem#1\endcsname}%
\let\auto@bib@innerbib\@empty
\bibitem [{\citenamefont {Rodnyi}(1997)}]{Rod97}%
  \BibitemOpen
  \bibfield  {author} {\bibinfo {author} {\bibfnamefont {P.~A.}\ \bibnamefont
  {Rodnyi}},\ }\href@noop {} {\emph {\bibinfo {title} {Physical processes in
  inorganic scintillators}}}\ (\bibinfo  {publisher} {{CRC} Press},\ \bibinfo
  {address} {Boca Raton},\ \bibinfo {year} {1997})\BibitemShut {NoStop}%
\bibitem [{\citenamefont {Knoll}(2010)}]{Kno10}%
  \BibitemOpen
  \bibfield  {author} {\bibinfo {author} {\bibfnamefont {G.~F.}\ \bibnamefont
  {Knoll}},\ }\href@noop {} {\emph {\bibinfo {title} {Radiation detection and
  measurement; 4th ed.}}}\ (\bibinfo  {publisher} {Wiley},\ \bibinfo {address}
  {New York, NY},\ \bibinfo {year} {2010})\BibitemShut {NoStop}%
\bibitem [{\citenamefont {Nelson}\ \emph {et~al.}(2011)\citenamefont {Nelson},
  \citenamefont {Gosnell},\ and\ \citenamefont {Knapp}}]{NelGosKna11}%
  \BibitemOpen
  \bibfield  {author} {\bibinfo {author} {\bibfnamefont {K.~E.}\ \bibnamefont
  {Nelson}}, \bibinfo {author} {\bibfnamefont {T.~B.}\ \bibnamefont {Gosnell}},
  \ and\ \bibinfo {author} {\bibfnamefont {D.~A.}\ \bibnamefont {Knapp}},\
  }\href {\doibase 10.1016/j.nima.2011.06.057} {\bibfield  {journal} {\bibinfo
  {journal} {Nucl. Instrum. Meth. A}\ }\textbf {\bibinfo {volume} {659}},\
  \bibinfo {pages} {207 } (\bibinfo {year} {2011})}\BibitemShut {NoStop}%
\bibitem [{\citenamefont {Dorenbos}(2010)}]{Dor10}%
  \BibitemOpen
  \bibfield  {author} {\bibinfo {author} {\bibfnamefont {P.}~\bibnamefont
  {Dorenbos}},\ }\href {\doibase 10.1109/TNS.2009.2031140} {\bibfield
  {journal} {\bibinfo  {journal} {IEEE Trans. Nucl. Sci.}\ }\textbf {\bibinfo
  {volume} {57}},\ \bibinfo {pages} {1162} (\bibinfo {year}
  {2010})}\BibitemShut {NoStop}%
\bibitem [{\citenamefont {Dorenbos}(2005)}]{Dor05}%
  \BibitemOpen
  \bibfield  {author} {\bibinfo {author} {\bibfnamefont {P.}~\bibnamefont
  {Dorenbos}},\ }\href {\doibase 10.1002/pssa.200460106} {\bibfield  {journal}
  {\bibinfo  {journal} {Phys. Status Solidi A}\ }\textbf {\bibinfo {volume}
  {202}},\ \bibinfo {pages} {195} (\bibinfo {year} {2005})}\BibitemShut
  {NoStop}%
\bibitem [{\citenamefont {Vasil'ev}(2008)}]{Vas08}%
  \BibitemOpen
  \bibfield  {author} {\bibinfo {author} {\bibfnamefont {A.~V.}\ \bibnamefont
  {Vasil'ev}},\ }\href {\doibase 10.1109/TNS.2007.914367} {\bibfield  {journal}
  {\bibinfo  {journal} {IEEE Trans. Nucl. Sci.}\ }\textbf {\bibinfo {volume}
  {55}},\ \bibinfo {pages} {1054} (\bibinfo {year} {2008})}\BibitemShut
  {NoStop}%
\bibitem [{\citenamefont {Kerisit}\ \emph {et~al.}(2009)\citenamefont
  {Kerisit}, \citenamefont {Rosso}, \citenamefont {Cannon}, \citenamefont
  {Gao},\ and\ \citenamefont {Xie}}]{KerRosCan09}%
  \BibitemOpen
  \bibfield  {author} {\bibinfo {author} {\bibfnamefont {S.}~\bibnamefont
  {Kerisit}}, \bibinfo {author} {\bibfnamefont {K.~M.}\ \bibnamefont {Rosso}},
  \bibinfo {author} {\bibfnamefont {B.~D.}\ \bibnamefont {Cannon}}, \bibinfo
  {author} {\bibfnamefont {F.}~\bibnamefont {Gao}}, \ and\ \bibinfo {author}
  {\bibfnamefont {Y.}~\bibnamefont {Xie}},\ }\href {\doibase 10.1063/1.3143786}
  {\bibfield  {journal} {\bibinfo  {journal} {J. Appl. Phys.}\ }\textbf
  {\bibinfo {volume} {105}},\ \bibinfo {pages} {114915} (\bibinfo {year}
  {2009})}\BibitemShut {NoStop}%
\bibitem [{\citenamefont {Bizarri}\ \emph {et~al.}(2009)\citenamefont
  {Bizarri}, \citenamefont {Moses}, \citenamefont {Singh}, \citenamefont
  {Vasil’ev},\ and\ \citenamefont {Williams}}]{BizMosSin09}%
  \BibitemOpen
  \bibfield  {author} {\bibinfo {author} {\bibfnamefont {G.}~\bibnamefont
  {Bizarri}}, \bibinfo {author} {\bibfnamefont {W.}~\bibnamefont {Moses}},
  \bibinfo {author} {\bibfnamefont {J.}~\bibnamefont {Singh}}, \bibinfo
  {author} {\bibfnamefont {A.}~\bibnamefont {Vasil’ev}}, \ and\ \bibinfo
  {author} {\bibfnamefont {R.}~\bibnamefont {Williams}},\ }\href {\doibase
  http://dx.doi.org/10.1016/j.jlumin.2008.12.024} {\bibfield  {journal}
  {\bibinfo  {journal} {J. Lumin.}\ }\textbf {\bibinfo {volume} {129}},\
  \bibinfo {pages} {1790 } (\bibinfo {year} {2009})}\BibitemShut {NoStop}%
\bibitem [{\citenamefont {Payne}\ \emph {et~al.}(2011)\citenamefont {Payne},
  \citenamefont {Moses}, \citenamefont {Sheets}, \citenamefont {Ahle},
  \citenamefont {Cherepy}, \citenamefont {Sturm}, \citenamefont {Dazeley},
  \citenamefont {Bizarri},\ and\ \citenamefont {Choong}}]{PayMosShe11}%
  \BibitemOpen
  \bibfield  {author} {\bibinfo {author} {\bibfnamefont {S.}~\bibnamefont
  {Payne}}, \bibinfo {author} {\bibfnamefont {W.~W.}\ \bibnamefont {Moses}},
  \bibinfo {author} {\bibfnamefont {S.}~\bibnamefont {Sheets}}, \bibinfo
  {author} {\bibfnamefont {L.}~\bibnamefont {Ahle}}, \bibinfo {author}
  {\bibfnamefont {N.}~\bibnamefont {Cherepy}}, \bibinfo {author} {\bibfnamefont
  {B.}~\bibnamefont {Sturm}}, \bibinfo {author} {\bibfnamefont
  {S.}~\bibnamefont {Dazeley}}, \bibinfo {author} {\bibfnamefont
  {G.}~\bibnamefont {Bizarri}}, \ and\ \bibinfo {author} {\bibfnamefont
  {W.-S.}\ \bibnamefont {Choong}},\ }\href {\doibase 10.1109/TNS.2011.2167687}
  {\bibfield  {journal} {\bibinfo  {journal} {{IEEE} Trans. Nucl. Sci.}\
  }\textbf {\bibinfo {volume} {58}},\ \bibinfo {pages} {3392} (\bibinfo {year}
  {2011})}\BibitemShut {NoStop}%
\bibitem [{\citenamefont {Yang}\ \emph {et~al.}(2012)\citenamefont {Yang},
  \citenamefont {Menge}, \citenamefont {Buzniak},\ and\ \citenamefont
  {Ouspenski}}]{YanMenBuz12}%
  \BibitemOpen
  \bibfield  {author} {\bibinfo {author} {\bibfnamefont {K.}~\bibnamefont
  {Yang}}, \bibinfo {author} {\bibfnamefont {P.}~\bibnamefont {Menge}},
  \bibinfo {author} {\bibfnamefont {J.}~\bibnamefont {Buzniak}}, \ and\
  \bibinfo {author} {\bibfnamefont {V.}~\bibnamefont {Ouspenski}},\ }in\ \href
  {\doibase 10.1109/NSSMIC.2012.6551113} {\emph {\bibinfo {booktitle} {Nuclear
  Science Symposium and Medical Imaging Conference (NSS/MIC), 2012 IEEE}}}\
  (\bibinfo {year} {2012})\ pp.\ \bibinfo {pages} {308--311}\BibitemShut
  {NoStop}%
\bibitem [{\citenamefont {Alekhin}\ \emph
  {et~al.}(2013{\natexlab{a}})\citenamefont {Alekhin}, \citenamefont {de~Haas},
  \citenamefont {Khodyuk}, \citenamefont {Kr\"amer}, \citenamefont {Menge},
  \citenamefont {Ouspenski},\ and\ \citenamefont {Dorenbos}}]{AleHaaKho13}%
  \BibitemOpen
  \bibfield  {author} {\bibinfo {author} {\bibfnamefont {M.~S.}\ \bibnamefont
  {Alekhin}}, \bibinfo {author} {\bibfnamefont {J.~T.~M.}\ \bibnamefont
  {de~Haas}}, \bibinfo {author} {\bibfnamefont {I.~V.}\ \bibnamefont
  {Khodyuk}}, \bibinfo {author} {\bibfnamefont {K.~W.}\ \bibnamefont
  {Kr\"amer}}, \bibinfo {author} {\bibfnamefont {P.~R.}\ \bibnamefont {Menge}},
  \bibinfo {author} {\bibfnamefont {V.}~\bibnamefont {Ouspenski}}, \ and\
  \bibinfo {author} {\bibfnamefont {P.}~\bibnamefont {Dorenbos}},\ }\href
  {\doibase doi:10.1063/1.4803440} {\bibfield  {journal} {\bibinfo  {journal}
  {Appl. Phys. Lett.}\ }\textbf {\bibinfo {volume} {102}},\ \bibinfo {pages}
  {161915} (\bibinfo {year} {2013}{\natexlab{a}})}\BibitemShut {NoStop}%
\bibitem [{\citenamefont {Dorenbos}\ \emph {et~al.}(2013)\citenamefont
  {Dorenbos}, \citenamefont {Alekhin}, \citenamefont {Khodyuk}, \citenamefont
  {de~Haas},\ and\ \citenamefont {Kr\"{a}mer}}]{DorAleKho13}%
  \BibitemOpen
  \bibfield  {author} {\bibinfo {author} {\bibfnamefont {P.}~\bibnamefont
  {Dorenbos}}, \bibinfo {author} {\bibfnamefont {M.}~\bibnamefont {Alekhin}},
  \bibinfo {author} {\bibfnamefont {I.~V.}\ \bibnamefont {Khodyuk}}, \bibinfo
  {author} {\bibfnamefont {J.~T.~M.}\ \bibnamefont {de~Haas}}, \ and\ \bibinfo
  {author} {\bibfnamefont {K.}~\bibnamefont {Kr\"{a}mer}}\ }(\bibinfo
  {publisher} {Presented at SCINT 2013, Shanghai, China},\ \bibinfo {year}
  {2013})\BibitemShut {NoStop}%
\bibitem [{\citenamefont {Alekhin}\ \emph {et~al.}(2014)\citenamefont
  {Alekhin}, \citenamefont {Weber}, \citenamefont {Kr\"amer},\ and\
  \citenamefont {Dorenbos}}]{AleWebKra14}%
  \BibitemOpen
  \bibfield  {author} {\bibinfo {author} {\bibfnamefont {M.~S.}\ \bibnamefont
  {Alekhin}}, \bibinfo {author} {\bibfnamefont {S.}~\bibnamefont {Weber}},
  \bibinfo {author} {\bibfnamefont {K.~W.}\ \bibnamefont {Kr\"amer}}, \ and\
  \bibinfo {author} {\bibfnamefont {P.}~\bibnamefont {Dorenbos}},\ }\href
  {\doibase 10.1016/j.jlumin.2013.08.019} {\bibfield  {journal} {\bibinfo
  {journal} {Journal of Luminescence}\ }\textbf {\bibinfo {volume} {145}},\
  \bibinfo {pages} {518} (\bibinfo {year} {2014})}\BibitemShut {NoStop}%
\bibitem [{\citenamefont {{\AA}berg}\ \emph {et~al.}(2014)\citenamefont
  {{\AA}berg}, \citenamefont {Sadigh}, \citenamefont {Schleife},\ and\
  \citenamefont {Erhart}}]{AbeSadSch14}%
  \BibitemOpen
  \bibfield  {author} {\bibinfo {author} {\bibfnamefont {D.}~\bibnamefont
  {{\AA}berg}}, \bibinfo {author} {\bibfnamefont {B.}~\bibnamefont {Sadigh}},
  \bibinfo {author} {\bibfnamefont {A.}~\bibnamefont {Schleife}}, \ and\
  \bibinfo {author} {\bibfnamefont {P.}~\bibnamefont {Erhart}},\ }\href
  {\doibase 10.1063/1.4880576} {\bibfield  {journal} {\bibinfo  {journal}
  {Applied Physics Letters}\ }\textbf {\bibinfo {volume} {104}},\ \bibinfo
  {pages} {211908} (\bibinfo {year} {2014})}\BibitemShut {NoStop}%
\bibitem [{\citenamefont {Bl\"ochl}(1994)}]{Blo94}%
  \BibitemOpen
  \bibfield  {author} {\bibinfo {author} {\bibfnamefont {P.~E.}\ \bibnamefont
  {Bl\"ochl}},\ }\href {\doibase 10.1103/PhysRevB.54.11169} {\bibfield
  {journal} {\bibinfo  {journal} {Phys. Rev. B}\ }\textbf {\bibinfo {volume}
  {50}},\ \bibinfo {pages} {17953 } (\bibinfo {year} {1994})}\BibitemShut
  {NoStop}%
\bibitem [{\citenamefont {Kresse}\ and\ \citenamefont
  {Joubert}(1999)}]{KreJou99}%
  \BibitemOpen
  \bibfield  {author} {\bibinfo {author} {\bibfnamefont {G.}~\bibnamefont
  {Kresse}}\ and\ \bibinfo {author} {\bibfnamefont {D.}~\bibnamefont
  {Joubert}},\ }\href {\doibase 10.1103/PhysRevB.59.1758} {\bibfield  {journal}
  {\bibinfo  {journal} {Phys. Rev. B}\ }\textbf {\bibinfo {volume} {59}},\
  \bibinfo {pages} {1758 } (\bibinfo {year} {1999})}\BibitemShut {NoStop}%
\bibitem [{\citenamefont {Kresse}\ and\ \citenamefont
  {Hafner}(1993)}]{KreHaf93}%
  \BibitemOpen
  \bibfield  {author} {\bibinfo {author} {\bibfnamefont {G.}~\bibnamefont
  {Kresse}}\ and\ \bibinfo {author} {\bibfnamefont {J.}~\bibnamefont
  {Hafner}},\ }\href {\doibase 10.1103/PhysRevB.47.558} {\bibfield  {journal}
  {\bibinfo  {journal} {Phys. Rev. B}\ }\textbf {\bibinfo {volume} {47}},\
  \bibinfo {pages} {558} (\bibinfo {year} {1993})}\BibitemShut {NoStop}%
\bibitem [{\citenamefont {Kresse}\ and\ \citenamefont
  {Hafner}(1994)}]{KreHaf94}%
  \BibitemOpen
  \bibfield  {author} {\bibinfo {author} {\bibfnamefont {G.}~\bibnamefont
  {Kresse}}\ and\ \bibinfo {author} {\bibfnamefont {J.}~\bibnamefont
  {Hafner}},\ }\href {\doibase 10.1103/PhysRevB.49.14251} {\bibfield  {journal}
  {\bibinfo  {journal} {Phys. Rev. B}\ }\textbf {\bibinfo {volume} {49}},\
  \bibinfo {pages} {14251} (\bibinfo {year} {1994})}\BibitemShut {NoStop}%
\bibitem [{\citenamefont {Kresse}\ and\ \citenamefont
  {Furthm\"uller}(1996{\natexlab{a}})}]{KreFur96a}%
  \BibitemOpen
  \bibfield  {author} {\bibinfo {author} {\bibfnamefont {G.}~\bibnamefont
  {Kresse}}\ and\ \bibinfo {author} {\bibfnamefont {J.}~\bibnamefont
  {Furthm\"uller}},\ }\href {\doibase 10.1103/PhysRevB.54.11169} {\bibfield
  {journal} {\bibinfo  {journal} {Phys. Rev. B}\ }\textbf {\bibinfo {volume}
  {54}},\ \bibinfo {pages} {11169} (\bibinfo {year}
  {1996}{\natexlab{a}})}\BibitemShut {NoStop}%
\bibitem [{\citenamefont {Kresse}\ and\ \citenamefont
  {Furthm\"uller}(1996{\natexlab{b}})}]{KreFur96b}%
  \BibitemOpen
  \bibfield  {author} {\bibinfo {author} {\bibfnamefont {G.}~\bibnamefont
  {Kresse}}\ and\ \bibinfo {author} {\bibfnamefont {J.}~\bibnamefont
  {Furthm\"uller}},\ }\href {\doibase 10.1016/0927-0256(96)00008-0} {\bibfield
  {journal} {\bibinfo  {journal} {Comput. Mater. Sci.}\ }\textbf {\bibinfo
  {volume} {6}},\ \bibinfo {pages} {15} (\bibinfo {year}
  {1996}{\natexlab{b}})}\BibitemShut {NoStop}%
\bibitem [{\citenamefont {Perdew}\ \emph {et~al.}(1996)\citenamefont {Perdew},
  \citenamefont {Burke},\ and\ \citenamefont {Ernzerhof}}]{PerBurErn96}%
  \BibitemOpen
  \bibfield  {author} {\bibinfo {author} {\bibfnamefont {J.~P.}\ \bibnamefont
  {Perdew}}, \bibinfo {author} {\bibfnamefont {K.}~\bibnamefont {Burke}}, \
  and\ \bibinfo {author} {\bibfnamefont {M.}~\bibnamefont {Ernzerhof}},\ }\href
  {\doibase 10.1103/PhysRevLett.77.3865} {\bibfield  {journal} {\bibinfo
  {journal} {Phys. Rev. Lett.}\ }\textbf {\bibinfo {volume} {77}},\ \bibinfo
  {pages} {3865} (\bibinfo {year} {1996})},\ \bibinfo {note} {erratum, {\it
  ibid.} \textbf{78}, 1396(E) (1997)}\BibitemShut {NoStop}%
\bibitem [{\citenamefont {Dudarev}\ \emph {et~al.}(1998)\citenamefont
  {Dudarev}, \citenamefont {Botton}, \citenamefont {Savrasov}, \citenamefont
  {Humphreys},\ and\ \citenamefont {Sutton}}]{DudBotSav98}%
  \BibitemOpen
  \bibfield  {author} {\bibinfo {author} {\bibfnamefont {S.~L.}\ \bibnamefont
  {Dudarev}}, \bibinfo {author} {\bibfnamefont {G.~A.}\ \bibnamefont {Botton}},
  \bibinfo {author} {\bibfnamefont {S.~Y.}\ \bibnamefont {Savrasov}}, \bibinfo
  {author} {\bibfnamefont {C.~J.}\ \bibnamefont {Humphreys}}, \ and\ \bibinfo
  {author} {\bibfnamefont {A.~P.}\ \bibnamefont {Sutton}},\ }\href@noop {}
  {\bibfield  {journal} {\bibinfo  {journal} {Phys. Rev. B}\ }\textbf {\bibinfo
  {volume} {57}},\ \bibinfo {pages} {1505} (\bibinfo {year}
  {1998})}\BibitemShut {NoStop}%
\bibitem [{\citenamefont {Czy\.{z}yk}\ and\ \citenamefont
  {Sawatzky}(1994)}]{CzySaw94}%
  \BibitemOpen
  \bibfield  {author} {\bibinfo {author} {\bibfnamefont {M.~T.}\ \bibnamefont
  {Czy\.{z}yk}}\ and\ \bibinfo {author} {\bibfnamefont {G.~A.}\ \bibnamefont
  {Sawatzky}},\ }\href
  {http://links.isiglobalnet2.com/gateway/Gateway.cgi?GWVersion=1&SrcAuth=KBib&SrcApp=KBib&KeyUT=A1994NR42300014}
  {\bibfield  {journal} {\bibinfo  {journal} {Phys. Rev. B}\ }\textbf {\bibinfo
  {volume} {49}},\ \bibinfo {pages} {14211} (\bibinfo {year}
  {1994})}\BibitemShut {NoStop}%
\bibitem [{\citenamefont {Canning}\ \emph {et~al.}(2011)\citenamefont
  {Canning}, \citenamefont {Chaudhry}, \citenamefont {Boutchko},\ and\
  \citenamefont {Gr{\o}nbech-Jensen}}]{CanChaBou11}%
  \BibitemOpen
  \bibfield  {author} {\bibinfo {author} {\bibfnamefont {A.}~\bibnamefont
  {Canning}}, \bibinfo {author} {\bibfnamefont {A.}~\bibnamefont {Chaudhry}},
  \bibinfo {author} {\bibfnamefont {R.}~\bibnamefont {Boutchko}}, \ and\
  \bibinfo {author} {\bibfnamefont {N.}~\bibnamefont {Gr{\o}nbech-Jensen}},\
  }\href {\doibase 10.1103/PhysRevB.83.125115} {\bibfield  {journal} {\bibinfo
  {journal} {Phys. Rev. B}\ }\textbf {\bibinfo {volume} {83}},\ \bibinfo
  {pages} {125115} (\bibinfo {year} {2011})}\BibitemShut {NoStop}%
\bibitem [{\citenamefont {{\AA}berg}\ \emph {et~al.}(2012)\citenamefont
  {{\AA}berg}, \citenamefont {Sadigh},\ and\ \citenamefont
  {Erhart}}]{AbeSadErh12}%
  \BibitemOpen
  \bibfield  {author} {\bibinfo {author} {\bibfnamefont {D.}~\bibnamefont
  {{\AA}berg}}, \bibinfo {author} {\bibfnamefont {B.}~\bibnamefont {Sadigh}}, \
  and\ \bibinfo {author} {\bibfnamefont {P.}~\bibnamefont {Erhart}},\ }\href
  {\doibase 10.1103/PhysRevB.85.125134} {\bibfield  {journal} {\bibinfo
  {journal} {Phys. Rev. B}\ }\textbf {\bibinfo {volume} {85}},\ \bibinfo
  {pages} {125134} (\bibinfo {year} {2012})}\BibitemShut {NoStop}%
\bibitem [{\citenamefont {Kr\"amer}\ \emph {et~al.}(1989)\citenamefont
  {Kr\"amer}, \citenamefont {Schleid}, \citenamefont {Schulze}, \citenamefont
  {urland},\ and\ \citenamefont {Meyer}}]{KraSchSch89}%
  \BibitemOpen
  \bibfield  {author} {\bibinfo {author} {\bibfnamefont {K.}~\bibnamefont
  {Kr\"amer}}, \bibinfo {author} {\bibfnamefont {T.}~\bibnamefont {Schleid}},
  \bibinfo {author} {\bibfnamefont {M.}~\bibnamefont {Schulze}}, \bibinfo
  {author} {\bibfnamefont {W.}~\bibnamefont {urland}}, \ and\ \bibinfo {author}
  {\bibfnamefont {G.}~\bibnamefont {Meyer}},\ }\href {\doibase
  10.1002/zaac.19895750109} {\bibfield  {journal} {\bibinfo  {journal}
  {Zeitschr. Anorg. Allg. Chemie}\ }\textbf {\bibinfo {volume} {575}},\
  \bibinfo {pages} {61} (\bibinfo {year} {1989})}\BibitemShut {NoStop}%
\bibitem [{\citenamefont {Chaudhry}\ \emph {et~al.}(2014)\citenamefont
  {Chaudhry}, \citenamefont {Boutchko}, \citenamefont {Chourou}, \citenamefont
  {Zhang}, \citenamefont {Gr{\o}nbech-Jensen},\ and\ \citenamefont
  {Canning}}]{ChaBouCho14}%
  \BibitemOpen
  \bibfield  {author} {\bibinfo {author} {\bibfnamefont {A.}~\bibnamefont
  {Chaudhry}}, \bibinfo {author} {\bibfnamefont {R.}~\bibnamefont {Boutchko}},
  \bibinfo {author} {\bibfnamefont {S.}~\bibnamefont {Chourou}}, \bibinfo
  {author} {\bibfnamefont {G.}~\bibnamefont {Zhang}}, \bibinfo {author}
  {\bibfnamefont {N.}~\bibnamefont {Gr{\o}nbech-Jensen}}, \ and\ \bibinfo {author}
  {\bibfnamefont {A.}~\bibnamefont {Canning}},\ }\href {\doibase
  10.1103/PhysRevB.89.155105} {\bibfield  {journal} {\bibinfo  {journal}
  {Physical Review B}\ }\textbf {\bibinfo {volume} {89}},\ \bibinfo {pages}
  {155105} (\bibinfo {year} {2014})}\BibitemShut {NoStop}%
\bibitem [{\citenamefont {Zhang}\ and\ \citenamefont
  {Northrup}(1991)}]{ZhaNor91}%
  \BibitemOpen
  \bibfield  {author} {\bibinfo {author} {\bibfnamefont {S.~B.}\ \bibnamefont
  {Zhang}}\ and\ \bibinfo {author} {\bibfnamefont {J.~E.}\ \bibnamefont
  {Northrup}},\ }\href@noop {} {\bibfield  {journal} {\bibinfo  {journal}
  {Phys. Rev. Lett.}\ }\textbf {\bibinfo {volume} {67}},\ \bibinfo {pages}
  {2339} (\bibinfo {year} {1991})}\BibitemShut {NoStop}%
\bibitem [{\citenamefont {Erhart}\ \emph {et~al.}(2010)\citenamefont {Erhart},
  \citenamefont {{\AA}berg},\ and\ \citenamefont {Lordi}}]{ErhAbeLor10}%
  \BibitemOpen
  \bibfield  {author} {\bibinfo {author} {\bibfnamefont {P.}~\bibnamefont
  {Erhart}}, \bibinfo {author} {\bibfnamefont {D.}~\bibnamefont {{\AA}berg}}, \
  and\ \bibinfo {author} {\bibfnamefont {V.}~\bibnamefont {Lordi}},\ }\href
  {\doibase 10.1103/PhysRevB.81.195216} {\bibfield  {journal} {\bibinfo
  {journal} {Phys. Rev. B}\ }\textbf {\bibinfo {volume} {81}},\ \bibinfo
  {pages} {195216} (\bibinfo {year} {2010})}\BibitemShut {NoStop}%
\bibitem [{\citenamefont {Lany}\ and\ \citenamefont {Zunger}(2008)}]{LanZun08}%
  \BibitemOpen
  \bibfield  {author} {\bibinfo {author} {\bibfnamefont {S.}~\bibnamefont
  {Lany}}\ and\ \bibinfo {author} {\bibfnamefont {A.}~\bibnamefont {Zunger}},\
  }\href {\doibase 10.1103/PhysRevB.78.235104} {\bibfield  {journal} {\bibinfo
  {journal} {Phys. Rev. B}\ }\textbf {\bibinfo {volume} {78}},\ \bibinfo
  {pages} {235104} (\bibinfo {year} {2008})}\BibitemShut {NoStop}%
\bibitem [{\citenamefont {Sadigh}\ \emph {et~al.}(2014)\citenamefont {Sadigh},
  \citenamefont {Erhart},\ and\ \citenamefont {{\AA}berg}}]{SadErhAbe14}%
  \BibitemOpen
  \bibfield  {author} {\bibinfo {author} {\bibfnamefont {B.}~\bibnamefont
  {Sadigh}}, \bibinfo {author} {\bibfnamefont {P.}~\bibnamefont {Erhart}}, \
  and\ \bibinfo {author} {\bibfnamefont {D.}~\bibnamefont {{\AA}berg}},\
  }\href@noop {} {\bibfield  {journal} {\bibinfo  {journal} {arXiv:1401.7137}\
  } (\bibinfo {year} {2014})}\BibitemShut {NoStop}%
\bibitem [{\citenamefont {Persson}\ \emph {et~al.}(2005)\citenamefont
  {Persson}, \citenamefont {Zhao}, \citenamefont {Lany},\ and\ \citenamefont
  {Zunger}}]{PerZhaLan05}%
  \BibitemOpen
  \bibfield  {author} {\bibinfo {author} {\bibfnamefont {C.}~\bibnamefont
  {Persson}}, \bibinfo {author} {\bibfnamefont {Y.-J.}\ \bibnamefont {Zhao}},
  \bibinfo {author} {\bibfnamefont {S.}~\bibnamefont {Lany}}, \ and\ \bibinfo
  {author} {\bibfnamefont {A.}~\bibnamefont {Zunger}},\ }\href@noop {}
  {\bibfield  {journal} {\bibinfo  {journal} {Phys. Rev. B}\ }\textbf {\bibinfo
  {volume} {72}},\ \bibinfo {pages} {035211} (\bibinfo {year}
  {2005})}\BibitemShut {NoStop}%
\bibitem [{\citenamefont {Erhart}\ and\ \citenamefont {Albe}(2007)}]{ErhAlb07}%
  \BibitemOpen
  \bibfield  {author} {\bibinfo {author} {\bibfnamefont {P.}~\bibnamefont
  {Erhart}}\ and\ \bibinfo {author} {\bibfnamefont {K.}~\bibnamefont {Albe}},\
  }\href {\doibase 10.1063/1.2801011} {\bibfield  {journal} {\bibinfo
  {journal} {J. Appl. Phys.}\ }\textbf {\bibinfo {volume} {102}},\ \bibinfo
  {pages} {084111} (\bibinfo {year} {2007})}\BibitemShut {NoStop}%
\bibitem [{\citenamefont {Erhart}\ and\ \citenamefont {Albe}(2008)}]{ErhAlb08}%
  \BibitemOpen
  \bibfield  {author} {\bibinfo {author} {\bibfnamefont {P.}~\bibnamefont
  {Erhart}}\ and\ \bibinfo {author} {\bibfnamefont {K.}~\bibnamefont {Albe}},\
  }\href {\doibase 10.1063/1.2956327} {\bibfield  {journal} {\bibinfo
  {journal} {J. Appl. Phys.}\ }\textbf {\bibinfo {volume} {104}},\ \bibinfo
  {pages} {044315} (\bibinfo {year} {2008})}\BibitemShut {NoStop}%
\bibitem [{\citenamefont {Hedin}(1965)}]{Hed65}%
  \BibitemOpen
  \bibfield  {author} {\bibinfo {author} {\bibfnamefont {L.}~\bibnamefont
  {Hedin}},\ }\href {\doibase 10.1103/PhysRev.139.A796} {\bibfield  {journal}
  {\bibinfo  {journal} {Phys. Rev.}\ }\textbf {\bibinfo {volume} {139}},\
  \bibinfo {pages} {A796} (\bibinfo {year} {1965})}\BibitemShut {NoStop}%
\bibitem [{\citenamefont {Hedin}\ and\ \citenamefont
  {Lundqvist}(1970)}]{HedLun70}%
  \BibitemOpen
  \bibfield  {author} {\bibinfo {author} {\bibfnamefont {L.}~\bibnamefont
  {Hedin}}\ and\ \bibinfo {author} {\bibfnamefont {S.}~\bibnamefont
  {Lundqvist}},\ }in\ \href
  {http://www.sciencedirect.com/science/article/pii/S0081194708606153} {\emph
  {\bibinfo {booktitle} {Solid State Physics}}},\ Vol.\ \bibinfo {volume}
  {Volume 23},\ \bibinfo {editor} {edited by\ \bibinfo {editor} {\bibfnamefont
  {D.~T.}\ \bibnamefont {Frederick~Seitz}}\ and\ \bibinfo {editor}
  {\bibfnamefont {H.}~\bibnamefont {Ehrenreich}}}\ (\bibinfo  {publisher}
  {Academic Press},\ \bibinfo {year} {1970})\ pp.\ \bibinfo {pages}
  {1--181}\BibitemShut {NoStop}%
\bibitem [{\citenamefont {Dorenbos}\ \emph {et~al.}(2006)\citenamefont
  {Dorenbos}, \citenamefont {van Loef}, \citenamefont {Vink}, \citenamefont
  {van~der Kolk}, \citenamefont {van Eijk}, \citenamefont {Kr\"amer},
  \citenamefont {G\"udel}, \citenamefont {Higgins},\ and\ \citenamefont
  {Shah}}]{DorLoeVin06}%
  \BibitemOpen
  \bibfield  {author} {\bibinfo {author} {\bibfnamefont {P.}~\bibnamefont
  {Dorenbos}}, \bibinfo {author} {\bibfnamefont {E.~V.~D.}\ \bibnamefont {van
  Loef}}, \bibinfo {author} {\bibfnamefont {A.~P.}\ \bibnamefont {Vink}},
  \bibinfo {author} {\bibfnamefont {E.}~\bibnamefont {van~der Kolk}}, \bibinfo
  {author} {\bibfnamefont {C.~W.~E.}\ \bibnamefont {van Eijk}}, \bibinfo
  {author} {\bibfnamefont {K.~W.}\ \bibnamefont {Kr\"amer}}, \bibinfo {author}
  {\bibfnamefont {H.~U.}\ \bibnamefont {G\"udel}}, \bibinfo {author}
  {\bibfnamefont {W.~M.}\ \bibnamefont {Higgins}}, \ and\ \bibinfo {author}
  {\bibfnamefont {K.~S.}\ \bibnamefont {Shah}},\ }\href {\doibase
  10.1016/j.jlumin.2005.04.016} {\bibfield  {journal} {\bibinfo  {journal} {J.
  Luminescence}\ }\textbf {\bibinfo {volume} {117}},\ \bibinfo {pages} {147}
  (\bibinfo {year} {2006})}\BibitemShut {NoStop}%
\bibitem [{Note1()}]{Note1}%
  \BibitemOpen
  \bibinfo {note} {We here adopt the convention that negative binding energies
  indicate attraction.}\BibitemShut {Stop}%
\bibitem [{\citenamefont {Andriessen}\ \emph {et~al.}(2007)\citenamefont
  {Andriessen}, \citenamefont {van~der Kolk},\ and\ \citenamefont
  {Dorenbos}}]{AndKolDor07}%
  \BibitemOpen
  \bibfield  {author} {\bibinfo {author} {\bibfnamefont {J.}~\bibnamefont
  {Andriessen}}, \bibinfo {author} {\bibfnamefont {E.}~\bibnamefont {van~der
  Kolk}}, \ and\ \bibinfo {author} {\bibfnamefont {P.}~\bibnamefont
  {Dorenbos}},\ }\href {\doibase 10.1103/PhysRevB.76.075124} {\bibfield
  {journal} {\bibinfo  {journal} {Phys. Rev. B}\ }\textbf {\bibinfo {volume}
  {76}},\ \bibinfo {pages} {075124} (\bibinfo {year} {2007})}\BibitemShut
  {NoStop}%
\bibitem [{\citenamefont {Alekhin}\ \emph
  {et~al.}(2013{\natexlab{b}})\citenamefont {Alekhin}, \citenamefont {Biner},
  \citenamefont {Kr\"amer},\ and\ \citenamefont {Dorenbos}}]{AleBinKra13}%
  \BibitemOpen
  \bibfield  {author} {\bibinfo {author} {\bibfnamefont {M.~S.}\ \bibnamefont
  {Alekhin}}, \bibinfo {author} {\bibfnamefont {D.~A.}\ \bibnamefont {Biner}},
  \bibinfo {author} {\bibfnamefont {K.~W.}\ \bibnamefont {Kr\"amer}}, \ and\
  \bibinfo {author} {\bibfnamefont {P.}~\bibnamefont {Dorenbos}},\ }\href
  {\doibase doi:10.1063/1.4810848} {\ \textbf {\bibinfo {volume} {113}},\
  \bibinfo {pages} {224904} (\bibinfo {year} {2013}{\natexlab{b}})}\BibitemShut
  {NoStop}%
\bibitem [{\citenamefont {Williams}\ \emph {et~al.}(2013)\citenamefont
  {Williams}, \citenamefont {Grim}, \citenamefont {Li}, \citenamefont {Ucer},
  \citenamefont {Bizarri}, \citenamefont {Kerisit}, \citenamefont {Gao},
  \citenamefont {Bhattacharya}, \citenamefont {Tupitsyn}, \citenamefont {Rowe},
  \citenamefont {Buliga},\ and\ \citenamefont {Burger}}]{WilGriLi13}%
  \BibitemOpen
  \bibfield  {author} {\bibinfo {author} {\bibfnamefont {R.~T.}\ \bibnamefont
  {Williams}}, \bibinfo {author} {\bibfnamefont {J.~Q.}\ \bibnamefont {Grim}},
  \bibinfo {author} {\bibfnamefont {Q.}~\bibnamefont {Li}}, \bibinfo {author}
  {\bibfnamefont {K.~B.}\ \bibnamefont {Ucer}}, \bibinfo {author}
  {\bibfnamefont {G.~A.}\ \bibnamefont {Bizarri}}, \bibinfo {author}
  {\bibfnamefont {S.}~\bibnamefont {Kerisit}}, \bibinfo {author} {\bibfnamefont
  {F.}~\bibnamefont {Gao}}, \bibinfo {author} {\bibfnamefont {P.}~\bibnamefont
  {Bhattacharya}}, \bibinfo {author} {\bibfnamefont {E.}~\bibnamefont
  {Tupitsyn}}, \bibinfo {author} {\bibfnamefont {E.}~\bibnamefont {Rowe}},
  \bibinfo {author} {\bibfnamefont {V.~M.}\ \bibnamefont {Buliga}}, \ and\
  \bibinfo {author} {\bibfnamefont {A.}~\bibnamefont {Burger}},\ }\href
  {\doibase 10.1117/12.2027716} {\bibfield  {journal} {\bibinfo  {journal}
  {Proc. SPIE}\ }\textbf {\bibinfo {volume} {8852}},\ \bibinfo {pages} {88520J}
  (\bibinfo {year} {2013})}\BibitemShut {NoStop}%
\bibitem [{\citenamefont {Grim}\ \emph {et~al.}(2013)\citenamefont {Grim},
  \citenamefont {Ucer}, \citenamefont {Burger}, \citenamefont {Bhattacharya},
  \citenamefont {Tupitsyn}, \citenamefont {Rowe}, \citenamefont {Buliga},
  \citenamefont {Trefilova}, \citenamefont {Gektin}, \citenamefont {Bizarri},
  \citenamefont {Moses},\ and\ \citenamefont {Williams}}]{GriUceBur13}%
  \BibitemOpen
  \bibfield  {author} {\bibinfo {author} {\bibfnamefont {J.~Q.}\ \bibnamefont
  {Grim}}, \bibinfo {author} {\bibfnamefont {K.~B.}\ \bibnamefont {Ucer}},
  \bibinfo {author} {\bibfnamefont {A.}~\bibnamefont {Burger}}, \bibinfo
  {author} {\bibfnamefont {P.}~\bibnamefont {Bhattacharya}}, \bibinfo {author}
  {\bibfnamefont {E.}~\bibnamefont {Tupitsyn}}, \bibinfo {author}
  {\bibfnamefont {E.}~\bibnamefont {Rowe}}, \bibinfo {author} {\bibfnamefont
  {V.~M.}\ \bibnamefont {Buliga}}, \bibinfo {author} {\bibfnamefont
  {L.}~\bibnamefont {Trefilova}}, \bibinfo {author} {\bibfnamefont
  {A.}~\bibnamefont {Gektin}}, \bibinfo {author} {\bibfnamefont {G.~A.}\
  \bibnamefont {Bizarri}}, \bibinfo {author} {\bibfnamefont {W.~W.}\
  \bibnamefont {Moses}}, \ and\ \bibinfo {author} {\bibfnamefont {R.~T.}\
  \bibnamefont {Williams}},\ }\href {\doibase 10.1103/PhysRevB.87.125117}
  {\bibfield  {journal} {\bibinfo  {journal} {Physical Review B}\ }\textbf
  {\bibinfo {volume} {87}},\ \bibinfo {pages} {125117} (\bibinfo {year}
  {2013})}\BibitemShut {NoStop}%
\bibitem [{\citenamefont {Ucer}\ \emph {et~al.}(2014)\citenamefont {Ucer},
  \citenamefont {Bizarri}, \citenamefont {Burger}, \citenamefont {Gektin},
  \citenamefont {Trefilova},\ and\ \citenamefont {Williams}}]{UceBizBur14}%
  \BibitemOpen
  \bibfield  {author} {\bibinfo {author} {\bibfnamefont {K.~B.}\ \bibnamefont
  {Ucer}}, \bibinfo {author} {\bibfnamefont {G.}~\bibnamefont {Bizarri}},
  \bibinfo {author} {\bibfnamefont {A.}~\bibnamefont {Burger}}, \bibinfo
  {author} {\bibfnamefont {A.}~\bibnamefont {Gektin}}, \bibinfo {author}
  {\bibfnamefont {L.}~\bibnamefont {Trefilova}}, \ and\ \bibinfo {author}
  {\bibfnamefont {R.~T.}\ \bibnamefont {Williams}},\ }\href {\doibase
  10.1103/PhysRevB.89.165112} {\bibfield  {journal} {\bibinfo  {journal} {Phys.
  Rev. B}\ }\textbf {\bibinfo {volume} {89}},\ \bibinfo {pages} {165112}
  (\bibinfo {year} {2014})}\BibitemShut {NoStop}%
\bibitem [{\citenamefont {{\AA}berg}\ \emph {et~al.}(2008)\citenamefont
  {{\AA}berg}, \citenamefont {Erhart}, \citenamefont {Williamson},\ and\
  \citenamefont {Lordi}}]{AbeErhWil08}%
  \BibitemOpen
  \bibfield  {author} {\bibinfo {author} {\bibfnamefont {D.}~\bibnamefont
  {{\AA}berg}}, \bibinfo {author} {\bibfnamefont {P.}~\bibnamefont {Erhart}},
  \bibinfo {author} {\bibfnamefont {A.~J.}\ \bibnamefont {Williamson}}, \ and\
  \bibinfo {author} {\bibfnamefont {V.}~\bibnamefont {Lordi}},\ }\href
  {\doibase 10.1103/PhysRevB.77.165206} {\bibfield  {journal} {\bibinfo
  {journal} {Phys. Rev. B}\ }\textbf {\bibinfo {volume} {77}},\ \bibinfo
  {pages} {165206} (\bibinfo {year} {2008})}\BibitemShut {NoStop}%
\bibitem [{\citenamefont {Qian}\ \emph {et~al.}(1988)\citenamefont {Qian},
  \citenamefont {Martin},\ and\ \citenamefont {Chadi}}]{QiaMarCha88}%
  \BibitemOpen
  \bibfield  {author} {\bibinfo {author} {\bibfnamefont {G.-X.}\ \bibnamefont
  {Qian}}, \bibinfo {author} {\bibfnamefont {R.~M.}\ \bibnamefont {Martin}}, \
  and\ \bibinfo {author} {\bibfnamefont {D.~J.}\ \bibnamefont {Chadi}},\
  }\href@noop {} {\bibfield  {journal} {\bibinfo  {journal} {Phys. Rev. B}\
  }\textbf {\bibinfo {volume} {38}},\ \bibinfo {pages} {7649} (\bibinfo {year}
  {1988})}\BibitemShut {NoStop}%
\bibitem [{\citenamefont {Komsa}\ \emph {et~al.}(2012)\citenamefont {Komsa},
  \citenamefont {Rantala},\ and\ \citenamefont {Pasquarello}}]{KomRanPas12}%
  \BibitemOpen
  \bibfield  {author} {\bibinfo {author} {\bibfnamefont {H.-P.}\ \bibnamefont
  {Komsa}}, \bibinfo {author} {\bibfnamefont {T.~T.}\ \bibnamefont {Rantala}},
  \ and\ \bibinfo {author} {\bibfnamefont {A.}~\bibnamefont {Pasquarello}},\
  }\href {\doibase 10.1103/PhysRevB.86.045112} {\bibfield  {journal} {\bibinfo
  {journal} {Phys. Rev. B}\ }\textbf {\bibinfo {volume} {86}},\ \bibinfo
  {pages} {045112} (\bibinfo {year} {2012})}\BibitemShut {NoStop}%
\bibitem [{\citenamefont {Castleton}\ and\ \citenamefont
  {Mirbt}(2004)}]{CasMir04}%
  \BibitemOpen
  \bibfield  {author} {\bibinfo {author} {\bibfnamefont {C.~W.~M.}\
  \bibnamefont {Castleton}}\ and\ \bibinfo {author} {\bibfnamefont
  {S.}~\bibnamefont {Mirbt}},\ }\href {\doibase 10.1103/PhysRevB.70.195202}
  {\bibfield  {journal} {\bibinfo  {journal} {Phys. Rev. B}\ }\textbf {\bibinfo
  {volume} {70}},\ \bibinfo {pages} {195202} (\bibinfo {year}
  {2004})}\BibitemShut {NoStop}%
\bibitem [{\citenamefont {Makov}\ and\ \citenamefont {Payne}(1995)}]{MakPay95}%
  \BibitemOpen
  \bibfield  {author} {\bibinfo {author} {\bibfnamefont {G.}~\bibnamefont
  {Makov}}\ and\ \bibinfo {author} {\bibfnamefont {M.~C.}\ \bibnamefont
  {Payne}},\ }\href {\doibase 10.1103/PhysRevB.51.4014} {\bibfield  {journal}
  {\bibinfo  {journal} {Phys. Rev. B}\ }\textbf {\bibinfo {volume} {51}},\
  \bibinfo {pages} {4014} (\bibinfo {year} {1995})}\BibitemShut {NoStop}%
\end{thebibliography}
\end{document}